\documentclass[usegraphicx,useAMS,usenatbib]{mn2e}
\usepackage{url}
\usepackage{lscape}    
\usepackage{multicol}  
\usepackage{verbatim}
\usepackage{multirow}
\usepackage{rotating}

\def \nhthree{\mbox{NH$_3$\,}}
\def \nhone{\mbox{NH$_3$\,(1,1)\,}}

\def \kms{\mbox{km\,s$^{-1}$}}
\def \vlsr{\mbox{V$_{\rm LSR}$\,}}
\def \vrms{\mbox{$v_{\rm rms}$\,}}
\def \cm2{\mbox{cm$^{-2}$}}
\def \cm3{\mbox{cm$^{-3}$}}

\def \tmb{\mbox{$T_{\rm mb}$\,}}
\def \ta{\mbox{$T_{\rm A}^{*}$\,}}
\def \tsys{\mbox{$T_{\rm SYS}$\,}}
\def \trms{\mbox{$T_{\rm rms}$\,}}
\def \trot{\mbox{$T_{\rm rot}$\,}}
\def \tkin{\mbox{$T_{\rm k}$\,}}

\def \hi{H{\sc i}\,}
\def \hii{H{\sc ii}\,}

\title[12\,mm line survey: W28 field TeV gamma-ray sources.] {12\,mm line survey of the dense molecular gas towards the W28 field TeV gamma-ray sources.}

\author[B. Nicholas et al.]
       {B. Nicholas$^{1}$\thanks{E-mail: brent.nicholas@adelaide.edu.au}, G. Rowell$^{1}$, M. G. Burton$^{2}$, A. Walsh$^{3}$, Y. Fukui$^{4}$, A. Kawamura$^{4}$, \newauthor 
         S. Longmore$^{5}$ and E. Keto$^{5}$.\\
$^{1}$School of Chemistry and Physics, Adelaide University, Adelaide, 5005,  Australia\\
$^{2}$School of Physics, University of New South Wales, Sydney, 2052, Australia\\
$^{3}$Centre for Astronomy, School of Engineering and Physical Sciences, James Cook University, Townsville, 4811, Australia\\
$^{4}$Department of Astrophysics, Nagoya University, Furocho, Chikusa-ku, Nagoya, Aichi, 464-8602, Japan\\
$^{5}$Harvard-Smithsonian Center for Astrophysics, 60 Garden Street, MS 78, Cambridge, MA 02138, USA}

\begin{document}

\date{\today}

\maketitle

\label{firstpage}

\begin{abstract}

We present 12\,mm Mopra observations of dense molecular gas towards the W28 supernova remnant (SNR) field. The focus is on the dense molecular gas towards the TeV gamma-ray sources detected by the H.E.S.S. telescopes, which likely trace the cosmic-rays from W28 and possibly other sources in the region. 
Using the NH$_{3}$ inversion transitions we reveal several dense cores inside the molecular clouds, the majority of which coincide with high-mass star formation and \hii regions, including the energetic ultra-compact \hii region G5.89-0.39. A key exception to this is the cloud north east of W28, which is well-known to be disrupted as evidenced by clusters of 1720\,MHz OH masers and broad CO line emission. Here we detect broad NH$_{3}$, up to the (9,9) transition, with linewidths up to 16\,\kms. This broad NH$_3$ emission spatially matches well with the TeV source HESS~J1801-233 and CO emission, and its velocity dispersion distribution suggests external disruption from the W28 SNR direction. 
Other lines are detected, such as HC$_{3}$N and HC$_{5}$N, H$_2$O masers, and many radio recombination lines, all of which are primarily found towards the southern high-mass star formation regions. These observations provide a new view onto the internal structures and dynamics of the dense molecular gas towards the W28 SNR field,
and in tandem with future higher resolution TeV gamma-ray observations   will offer the chance to probe the transport of cosmic-rays into molecular clouds.

\end{abstract}


\section{Introduction}

\begin{figure}
  \centering
  \includegraphics[width=0.48\textwidth]{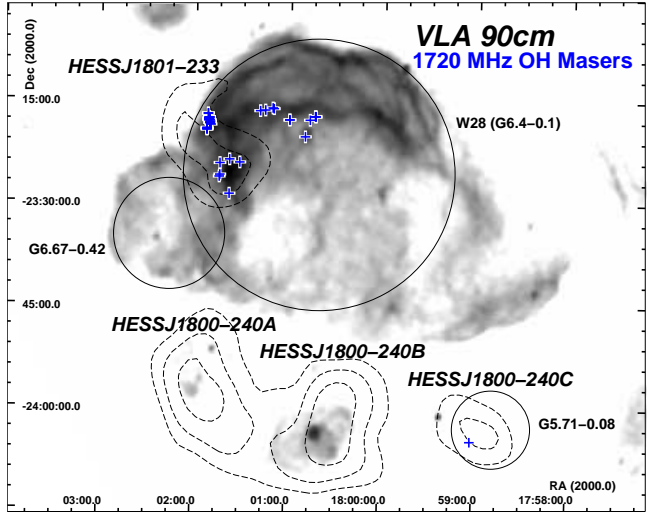}
  \caption{ VLA 90\,cm image (log scale up to 0.84\,Jy/beam) from \citet{brogan2006} of the W28 field with H.E.S.S. TeV emission (significance contours; black dashed). Boundaries of catalogued SNRs (black solid circles) and 1720\,MHz\,OH masers from \citet{claussen} and \citet{hewitt} (blue $+$'s) are also indicated.}
  \label{fig:vla}
\end{figure}

W28 (G6.4-0.1) is an old-age ($> 10^{4}$\,yr; \citealt{kaspi}), mixed morphology supernova remnant (SNR) spanning 50$^\prime \times 45^\prime$ with a distance estimated to be 
in the range 1.8 to 3.3\,kpc (e.g. \citealt{goudis,lozinskaya}). The SNR exhibits non-thermal radio emission and thermal X-rays \citep{dubner,rho2002}, and more recently, gamma-ray sources at TeV (10$^{12}$\,eV) \citep{hess_w28} and GeV (10$^{9}$\,eV) \citep{agile,fermi_w28} energies have been discovered by H.E.S.S., {\em AGILE}, and {\em Fermi}-LAT telescopes respectively, 
pointing to high energy particles in the region. 
$^{12}$CO\,(1--0), $^{12}$CO\,(2--1) and $^{12}$CO\,(3--2) surveys reveal massive molecular clouds to the north east (NE) and to the south (S) of the SNR \citep{arikawa,reach,torres,hess_w28,nanten21}. Most of the CO emission appears centred at a local standard of rest velocity \vlsr similar to that inferred for W28 \vlsr $\sim$7\,\kms (or $\sim 2$\,kpc) based on \hi studies \citep{velazquez}. \citet{torres} has argued that W28 has disrupted much of this CO gas, giving rise to its relatively broad velocity distribution. Notably, the NE region contains a rich concentration of 1720\,MHz OH masers \citep{frail,claussen} (with \vlsr in the range 5 to 15\,\kms), and near-IR rovibrational H$_2$ emission \citep{reach2000,neufeld,marquez-lugo}, all indicating shocked gas which likely results from a SNR shock interaction with the NE molecular cloud.
The southern region contains several \hii regions (G6.225-0.569, G6.1-0.6) including the ultra-compact (UC)-\hii region W28~A2 (G5.89-0.39), all indicating high-mass star formation. Additional SNRs have also been catalogued towards the W28 region, namely, G6.67-0.42 by \citet{yusef}, and G5.71-0.08 by \citet{brogan2006}. The $\sim$5~arcmin resolution H.E.S.S. TeV gamma-ray emission is resolved into four sources. HESS~J1800-233 is situated towards the NE region 
where a SNR shock is known to interact with a molecular cloud,
while a group of three TeV peaks are found towards the south coinciding with the \hii regions (HESS~J1800-240A and B) and the SNR candiate G5.71-0.08 (HESS~J1800-240C). Interestingly, a recently detected 1720\,MHz OH maser \citep{hewitt} towards G5.71-0.08, with \vlsr$=8$\,\kms, may also suggest HESS~J1800-240C is tracing a SNR/molecular cloud interaction. 
At lower angular resolution ($\sim$5 to 20~arcmin), two {\em Fermi}-LAT GeV sources, 1FGL~J1800.5-2359c and 1FGL~J1801.3-2322c (also detected by the {\em AGILE} detector) 
appear as counterparts to HESS~J1800-233 and HESS~J1800-240B respectively. 
Figure~\ref{fig:vla} compares the H.E.S.S. TeV emission with the 90\,cm radio (VLA) image from \citet{brogan2006}, highlighting the prominent W28 SNR emission which peaks towards the NE interaction region, and G5.89-0.89 to the south.

The TeV and GeV gamma-ray emission in the W28 region is spatially well-matched with the molecular clouds, and represents the best such match outside of the central molecular zone (CMZ) towards the Galactic Centre region \citep{hess_diffuse}. This, coupled with the old-age of W28, which would reduce any potential gamma-ray emission from accelerated electrons, suggests the gamma-ray emission results from collisions of cosmic-ray (CR) protons and nuclei with the molecular gas \citep{hess_w28,fujita}. 
W28 is a prominent member of a growing list of SNRs linked to TeV/GeV gamma-ray emission spatially matched with molecular clouds. This list includes HESS~J1745-290/SNR~G359.1-0.5 \citep{hess_1745}, HESS~J1714-385/CTB~37A \citep{hess_ctb37a}, HESS~J1923+141/SNR~G49.2-0.7 \citep{feinstein}, and IC~443 \citep{ic443_magic,ic443_veritas}. All of these SNRs exhibit 1720\,MHz OH masers as for W28, and appear to be mature SNRs (age $>10^{4}$\,yr) whereby any accelerated CRs would have begun to escape into the surrounding interstellar medium. 

In the W28 field, the obvious source of CRs is the W28 SNR given its prominence in many wavebands, but the other SNRs in the region, all with unknown distances and ages, may contribute to CR acceleration. Moreover, the extensive star formation and \hii regions associated with the southern molecular clouds, in particular the energetic UC-\hii region G5.89-0.39 may also contribute CRs based on recent discussion of protostellar particle acceleration \citep{araudo}. Some insight into where CRs are coming from can be obtained by looking at cloud density and emission line profiles in order to trace the presence and directionality of shocks.

Additionally, \citet{gabici} showed that the energy and magnetic field dependent diffusion of CRs can lead to a hardening of the TeV gamma-ray emission spectrum as one looks inward towards dense molecular cloud cores (which typically span scale of a few arcminutes). Thus, spatial knowledge of molecular cloud density structures and cores towards TeV gamma-ray sources is a key step in probing the long sought-after diffusion properties of CRs.

Since the abundant $^{12}$CO molecular cloud tracer, with critical hydrogen density $\sim 10^{2-3}$\,\cm3, rapidly becomes optically thick towards molecular cloud clumps and cores, understanding of 
molecular cloud density profiles and internal dynamics can be impaired. 
Ideal tracers of dense gas such as NH$_{3}$ (ammonia), CS, and HC$_{3}$N are widely used due to their lower abundance (a factor $\sim10^{-5}\times$CO) 
and higher critical densities $\sim 10^{4-5}$\,\cm3, which gives them a much lower optical thickness in dense gas. 
NH$_{3}$ is exceptionally useful since a single receiver at 23-25\,GHz can detect a large number of NH$_{3}$ inversion transitions 
which are produced over a narrow bandwidth. Through satellite and hyperfine structure, these inversion transitions also allow the optical depth, and hence gas temperature and mass to be strongly constrained. 
Another desirable property of NH$_{3}$ is that the different inversion transitions cover a wide range of excitation conditions. NH$_{3}$ has therefore been detected in many astrophysical environments 
such as dense quiescent/cold gas and both warm to hot gas in low and high mass star formation regions. Indeed, virtually any region containing dense molecular material can be studied 
with an appropriate NH$_{3}$ transition \citep{hotownes1983}.
NH$_{3}$ is also known to exist in the coldest regions of molecular clouds depleting less rapidly from the gas phase compared to other common gas tracers, such as CO, which tends to freeze out onto dust grains \citep{bergin2006}.

As the next step in probing the dense cores and dynamics of the W28 field molecular clouds, we have used the Mopra 22m single dish radio telescope in a 12\,mm survey and single position-switched pointing covering the key inversion transitions of NH$_{3}$ and several other 12\,mm lines tracing high-mass star formation including H$_2$O masers, HC$_3$N and CH$_3$OH.

\section{Mopra Observations and Data Reduction}
\label{sec:obs}

Observations were carried out on the Mopra radio telescope in May/June of 2008 and April of 2009 employing the Mopra spectrometer (MOPS) in zoom-mode. 

Mopra is a 22\,m single-dish radio telescope ($31^\circ 16^\prime 04^{\prime\prime}$S, $149^\circ 05^\prime 59^{\prime\prime}$E, 866m a.s.l.) located $\sim$450\,km north west of Sydney, Australia. The 12\,mm receiver operating in the frequency range of 16-27.5\,GHz, coupled with the UNSW Mopra wide-bandwidth spectrometer (MOPS), allows an instantaneous 8\,GHz bandwidth. This gives Mopra the ability to cover most of the 12\,mm band and simultaneously observe many spectral lines. The zoom-mode of MOPS allows observations from up to 16 windows simultaneously, where each window is 137.5\,MHz wide and contains 4096 channels in each of two polarisations. At 12\,mm this gives MOPS an effective bandwidth of $\sim$\,1800\,\kms\, with resolution $\sim$\,0.41\,\kms. 
Within this band the Mopra beam FWHM varies from 2.4$^\prime$ (19\,GHz) to 1.7$^\prime$ (27\,GHz) \citep{mopra_beam}. Listed in Table~\ref{tab:lines} are some of the lines which are simultaneously within the bandpass in our configuration.

\begin{table}
\centering
\caption{Molecular lines and the corresponding rest frequencies which are detectable by the MOPS spectrometer in the configuration used. The final two columns indicate whether we detect the line in our maps or a deep pointing. Methanol (CH$_{3}$OH) masers are listed as type I or type II.\label{tab:lines}}
\normalsize
\begin{tabular}{llcc}
\hline
Molecular Line & Frequency & Detected & Detected \\
\multicolumn{1}{c}{Name} & \multicolumn{1}{c}{(MHz)} & Map & Deep Spectra\\
\hline
H69$\alpha$    	& 19591.11   	& Yes 	& Yes\\
CH$_{3}$OH(II)	& 19967.396  	& -- 		& --\\
H86$\beta$ 	& 19978.17	& --		& Yes\\
H98$\gamma$	& 20036.32	& --		& Yes\\
NH$_{3}$\,(8,6)	& 20719.221  	& -- 		& --\\
NH$_{3}$\,(9,7) 	& 20735.452  	& -- 		& --\\
C$_{6}$H      	& 20792.872  	& -- 		& --\\
NH$_{3}$\,(7,5) 	& 20804.83   	& -- 		& --\\
NH$_{3}$\,(11,9)	& 21070.739  	& -- 		& --\\
NH$_{3}$\,(4,1) 	& 21134.311  	& -- 		& --\\
H83$\beta$ 	& 22196.47	& --		& Yes\\
H$_{2}$O Maser	& 22235.253  	& Yes 	& Yes\\
C$_{2}$S       	& 22344.030  	& -- 		& --\\
H82$\beta$ 	& 23008.61	& --		& Yes\\
NH$_{3}$\,(2,1) 	& 23098.819  	& -- 		& --\\
CH$_{3}$OH(II)	& 23121.024  	& -- 		& --\\
H65$\alpha$    	& 23404.28   	& Yes 	& Yes\\
CH$_{3}$OH     	& 23444.778 	& -- 		& --\\
NH$_{3}$\,(1,1) 	& 23694.4709 	& Yes 	& Yes\\
NH$_{3}$\,(2,2) 	& 23722.6336 	& Yes 	& Yes\\
H81$\beta$ 	& 23860.87	& --		& Yes\\
NH$_{3}$\,(3,3) 	& 23870.1296 	& Yes 	& Yes\\
CH$_{3}$OH(I) 	& 24928.715  	& -- 		& --\\
CH$_{3}$OH(I) 	& 24933.468  	& -- 		& --\\
CH$_{3}$OH(I) 	& 24934.382  	& -- 		& --\\
CH$_{3}$OH(I) 	& 24959.079  	& -- 		& --\\
CH$_{3}$OH(I) 	& 25018.123  	& -- 		& --\\
NH$_{3}$\,(6,6) 	& 25056.025  	& Yes	& Yes\\
CH$_{3}$OH(I) 	& 25124.872  	& -- 		& --\\
HC$_{5}$N(10--9) & 26626.533  	& Yes	& Yes\\
H89$\gamma$	& 26630.71	& --		& Yes\\
H78$\beta$ 	& 26684.34	& -- 		& Yes\\
CH$_{3}$OH(I) 	& 26847.205  	& -- 		& --\\
H62$\alpha$    	& 26939.17   	& Yes	& Yes\\
HC$_{3}$N(3--2)	& 27294.078  	& Yes 	& Yes\\
CH$_{3}$OH(I) 	& 27472.501  	& -- 		& --\\
NH$_{3}$\,(9,9) 	& 27477.943  	& -- 		& Yes\\
\hline
\end{tabular}
\end{table}

On-the-fly Mapping (OTF) observations were conducted in May of 2008, and consisted of four regions, which are referred as A (25$^{\prime}$\,$\times$\,$25^{\prime}$), B (20$^{\prime}$\,$\times$\,$20^{\prime}$) and C (15$^{\prime}$\,$\times$\,$15^{\prime}$) to cover the TeV emission peaks from HESS~J1800-240 \citep{hess_w28} and map D (20$^{\prime}$\,$\times$\,$20^{\prime}$) to cover the TeV emission from HESS~1801-233 \citep{hess_w28}. We mapped each region twice, scanning once in right ascension and once in declination in order to reduce noise levels and to eliminate artificial stripes that can be introduced when only one scanning direction is used. It also allows us to check for artefacts that may occur in one scan, but not the other. For reference, the mapped regions are indicated as dashed boxes in Figure~\ref{fig:peakpixmapNH311}.

The OTF mapping parameters we used are similar to those used in the the H$_2$O Southern Galactic Plane Survey (HOPS), a 12\,mm study of the Galactic Plane \citep{hops_pilot}. Since HOPS also covered the W28 region, we have also included HOPS data in our mapping analysis, improving our exposure in the mapped areas B, C \& D by a factor of two. Map~A extended beyond the Galactic latitude limit of HOPS ($b=-0.5^\circ$) and so only partial overlap ($\sim$25\%) exists.  

Based on mapping results and prior knowledge of the regions under consideration, follow-up single pointing position-switched deep spectra were performed in June 2008 and in April of 2009, to provide high sensitivity to yield accurate measurement of the \nhone satellite lines which are necessary to determine the gas temperature and density. As we show shortly, several regions were found to exhibit \nhone and higher transitions with satellite lines apparent in the (1,1) spectra. Regions of bright \nhone emission from each map were targeted in these deep spectra. In total there were 10 regions which were selected for follow-up spectra which consisted of 1920\,s (32\,min) of ON source time.

Data were reduced using the ATNF packages {\tt~livedata}, {\tt~gridzilla}, {\tt~ASAP} and {\tt Miriad}.\footnote{See \url{http://www.atnf.csiro.au/computing/software/} for more information on these data reduction packages.} 
For mapping, {\tt livedata} was used to perform a bandpass calibration for each row, using the preceding off scan as a reference and applied a 1$^{\rm st}$ order polynomial fit (i.e. linear) to the baseline. {\tt Gridzilla} re-gridded and combined all data from all mapping scans onto a single data cube with pixels 15$^{\prime\prime}\times15^{\prime\prime}\times$0.43\,\kms\,($x,y,z$). The mapping data are also weighted according to the relevant \tsys, Gaussian-smoothed (2$^\prime$ FWHM and 5$^\prime$ cut-off radius) based on the Mopra beam FWHM $\theta_{\rm mb}=2^\prime$ appropriate for 
the NH$_3$ lines we detected, and pixel masked to remove noisy edge pixels. Analysis of position-switched deep pointings employed {\tt ASAP} with time-averaging, weighting by the relevant \tsys and baseline subtracted using a linear fit after masking of the 15 channels at each bandpass edge. In both mapping and position-switched data, the antenna temperature \ta (corrected for atmospheric attenuation and rearward loss) is converted to the main beam brightness temperature \tmb, such that \tmb = \ta $\eta_{\rm mb}$ where  $\eta_{\rm mb}$ is the main beam efficiency. 
Based on the frequencies of the detected NH$_3$ lines ($\sim$24\,GHz) we assume $\eta_{\rm mb}=0.6$ following \citet{mopra_beam}. This data reduction procedure yields an RMS error in \tmb of \trms $\sim0.05$\,K per channel for mapping data with HOPS overlap and \trms $\sim0.08$\,K per channel for mapping data without. As a result of their increased exposure, position-switched observations achieve a \trms of $\sim0.02$\,K per channel. 

\section{Results Overview}
\label{sec:overview}

\begin{figure}
  \includegraphics[width=\columnwidth]{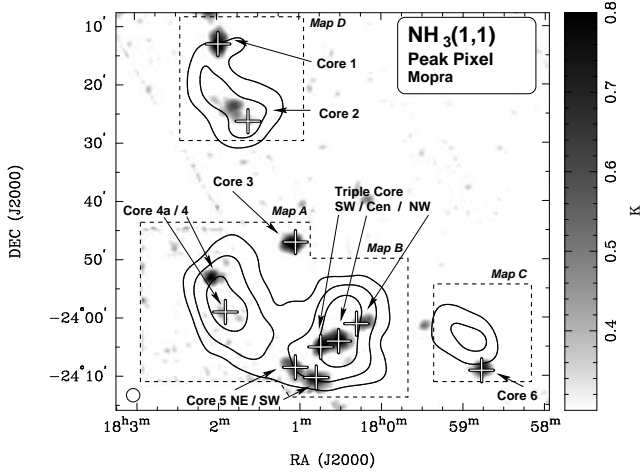}
  \caption{NH$_{3}$\,(1,1) peak pixel emission [\tmb] indicating the four mapped regions A,B,C,D in relation to the H.E.S.S. TeV emission (black contours 4, 5 and 6\,$\sigma$ levels). HOPS data are also included. Black/white $+$ indicate the locations of the position-switched deep spectra summarised in Table~\ref{tab:corePos}.}
  \label{fig:peakpixmapNH311}
\end{figure}

Table~\ref{tab:lines} lists the lines detected in our mapping and position-switched deep spectra observations.
In mapping data, the so called peak-pixel map for \nhone emission is shown in Figure~\ref{fig:peakpixmapNH311} along with the locations of position-switched deep pointings.
\begin{table}
\centering
\caption{Coordinates of the Mopra 12\,mm position-switched deep spectra observations and
  identification with the nearest TeV gamma-ray source. The position for Core~4 however, is taken from
  the mapping results since a position-switched observation was not taken.
  \label{tab:corePos}}
  \normalsize
  \begin{tabular}{lll}
  \hline
  Core Name & RA (J2000) & Dec. (J2000) \\
            & hh:mm:ss & dd:mm:ss \\
  \hline
  \multicolumn{3}{c}{Map D -- HESS J1801-233}\\
  Core 1			& 18:01:59  & -23:13:04\\	
  Core 2			& 18:01:37  & -23:26:21\\
  \multicolumn{3}{c}{Map A -- HESS J1800-240A}\\
  Core 3			& 18:01:03  	& -23:47:08\\	
  Core 4 			& 18:02:04      & -23:53:04\\          	
  Core 4a			& 18:01:55	& -23:59:05\\	
  \multicolumn{3}{c}{Map B -- HESS J1800-240B}\\
  Core 5 SW         		& 18:00:48	& -24:10:23\\	
  Core 5 NE		        & 18:01:03	& -24:08:38\\	
  Triple Core SE		& 18:00:45	& -24:05:08\\	
  Triple Core Central   	& 18:00:31	& -24:04:09\\	
  Triple Core NW		& 18:00:18	& -24:01:09\\	
  \multicolumn{3}{c}{Map C -- HESS J1800-240C}\\
  Core 6			& 17:58:46	& -24:09:10\\	
  \hline
\end{tabular}
\end{table}
Peak-pixel maps highlight only the brightest pixel along the velocity axis ($z$ axis) and serve as a useful way to search for pointlike and moderately extended
features. 

Of the 29 molecular lines listed in Table~\ref{tab:lines} within the MOPS bandwidth, about half were detected. From the mapping observations we detected H$_{2}$O, NH$_{3}$\,(1,1), NH$_{3}$\,(2,2), NH$_{3}$\,(3,3), NH$_{3}$\,(6,6), HC$_{3}$N(3--2),  HC$_{5}$N(10--9), H69$\alpha$, H65$\alpha$ and H62$\alpha$, with the criterion for detection being a 3 \trms line peak signal. Since position-switched deep spectra observations were more sensitive than mapping, several additional lines were revealed as can be seen in  Table~\ref{tab:lines}. Maps of position-velocity (PV), integrated intensity, and spectra for all detected lines can be found in the online appendix. 

\subsection{NH$_3$}
\begin{figure*}
  \centering
  \includegraphics[width=0.9\textwidth]{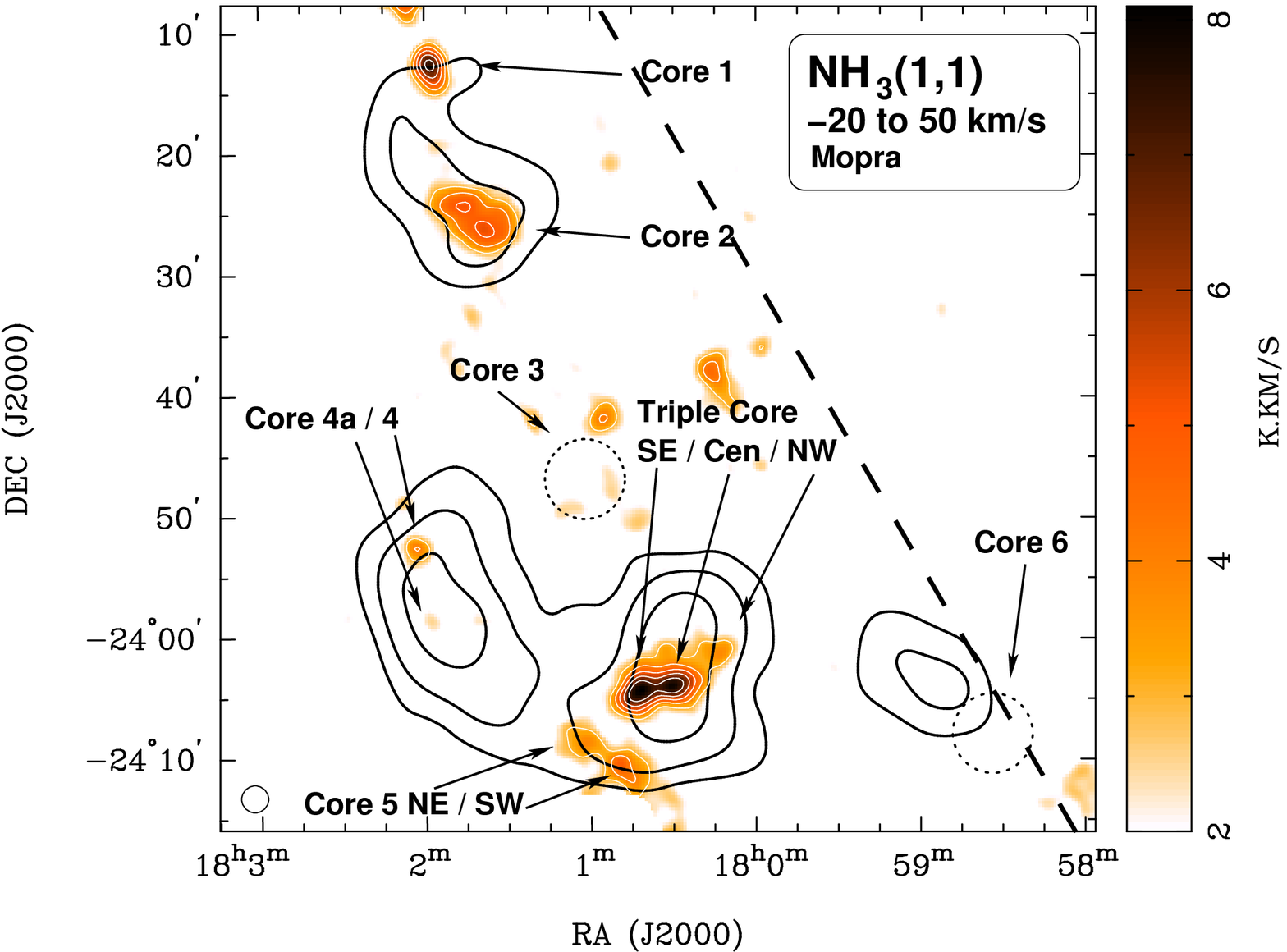}
  \includegraphics[width=0.9\textwidth]{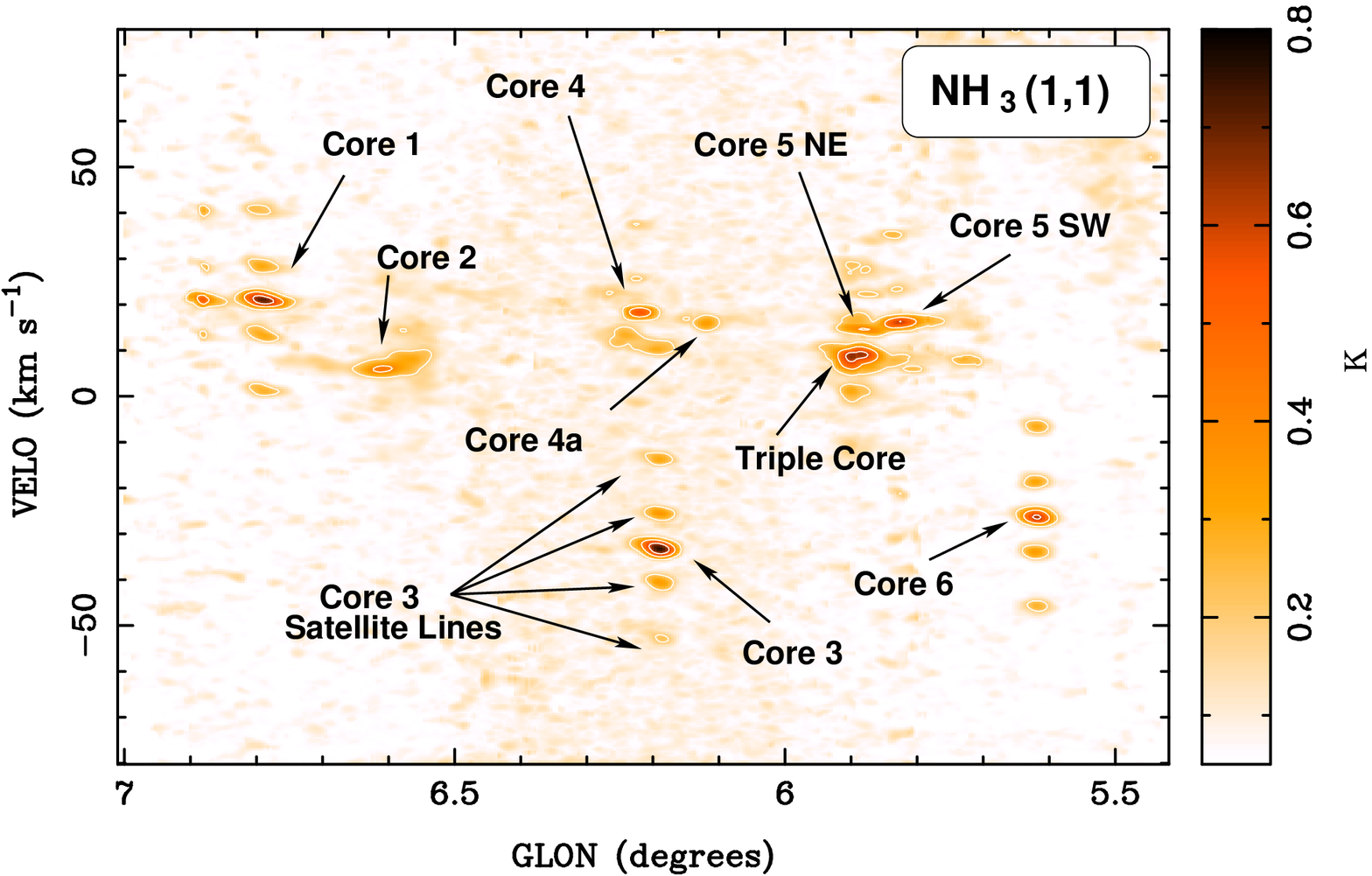}
  \caption[NH$_{3}$\,(1,1) Cores]{{\bf Top:} Integrated intensity [\tmb] map of Mopra NH$_{3}$\,(1,1) emission integrated from 
    -20 to 50\,\kms\, to encompass the various (1,1) cores towards the TeV emission (white contours are 3, 4, 5, 6, 7 K\,\kms). 
    Identified cores are labeled as described in the text. Dashed line indicates the Galactic Plane, dashed circles indicate the location of Cores~3 and 6 which have centre 
    velocities outside the integration range. The Mopra beam (2$^\prime$ FWHM) is indicated in the bottom left corner.
    {\bf Bottom:} Position-Velocity (PV) map of the peak pixel NH$_3$\,(1,1) emission (white contours 0.2, 0.4, and 0.6\,K) 
    indicating its velocity location and spread. For most cores the
    NH$_3$\,(1,1) satellite peaks (e.g. labeled for Core~3) may be seen as four peaks surrounding the central one along the velocity axis.}
  \label{fig:11cores}
\end{figure*}

\begin{figure*}
  \centering
  \includegraphics[width=0.9\textwidth]{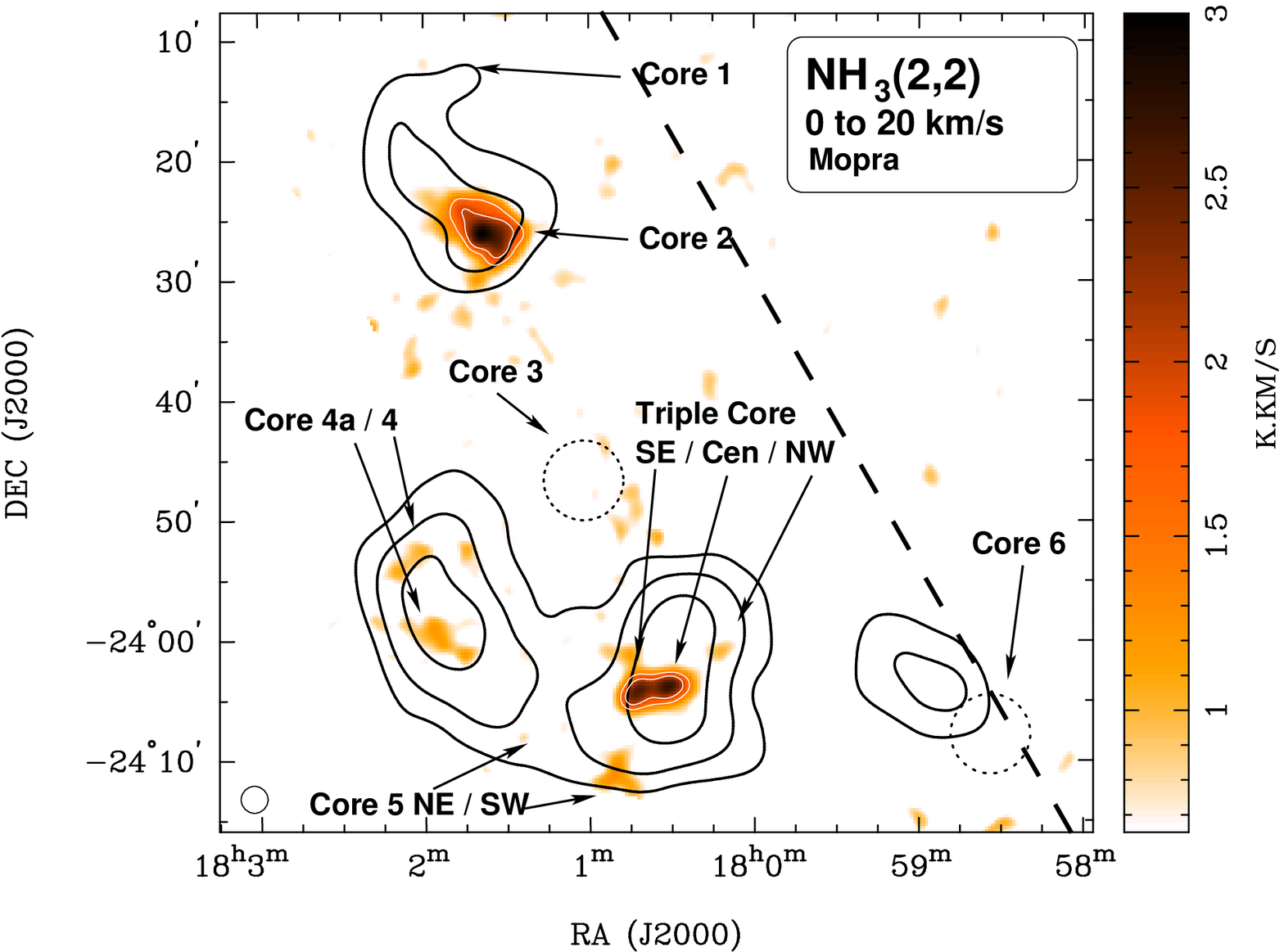}
  \includegraphics[width=0.9\textwidth]{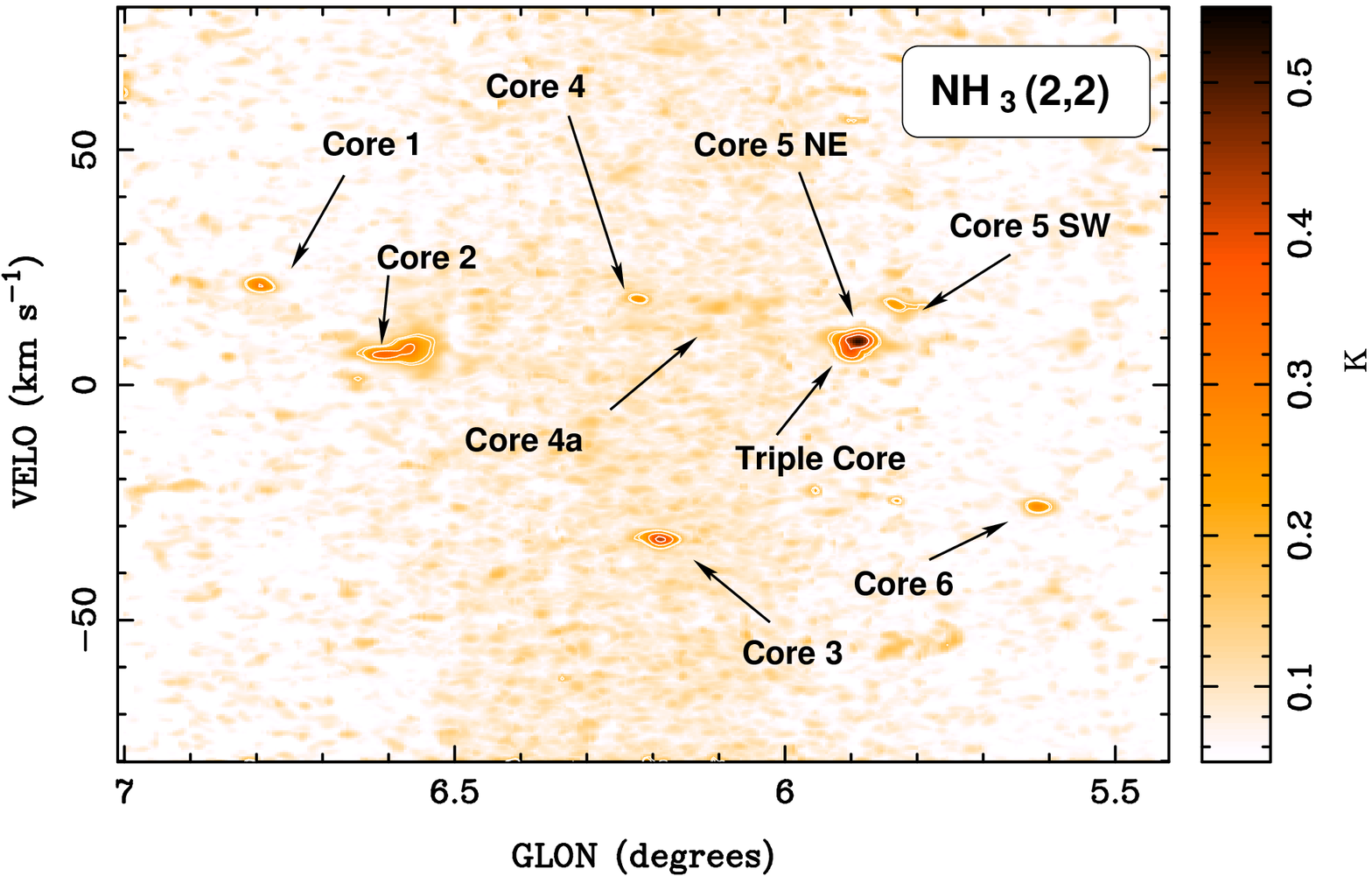}
  \caption[NH$_{3}$\,(2,2) Cores]{As for Figure~\ref{fig:11cores} but showing NH$_{3}$\,(2,2) integrated over 0 to 20\,\kms\, (white contours are 1.5, 2, 3 K\,\kms)
    and PV plot (white contours 0.18, 0.2, 0.3, and 0.4\,K).}
  \label{fig:22cores}
\end{figure*}

\begin{figure*}
  \centering
  \includegraphics[width=0.9\textwidth]{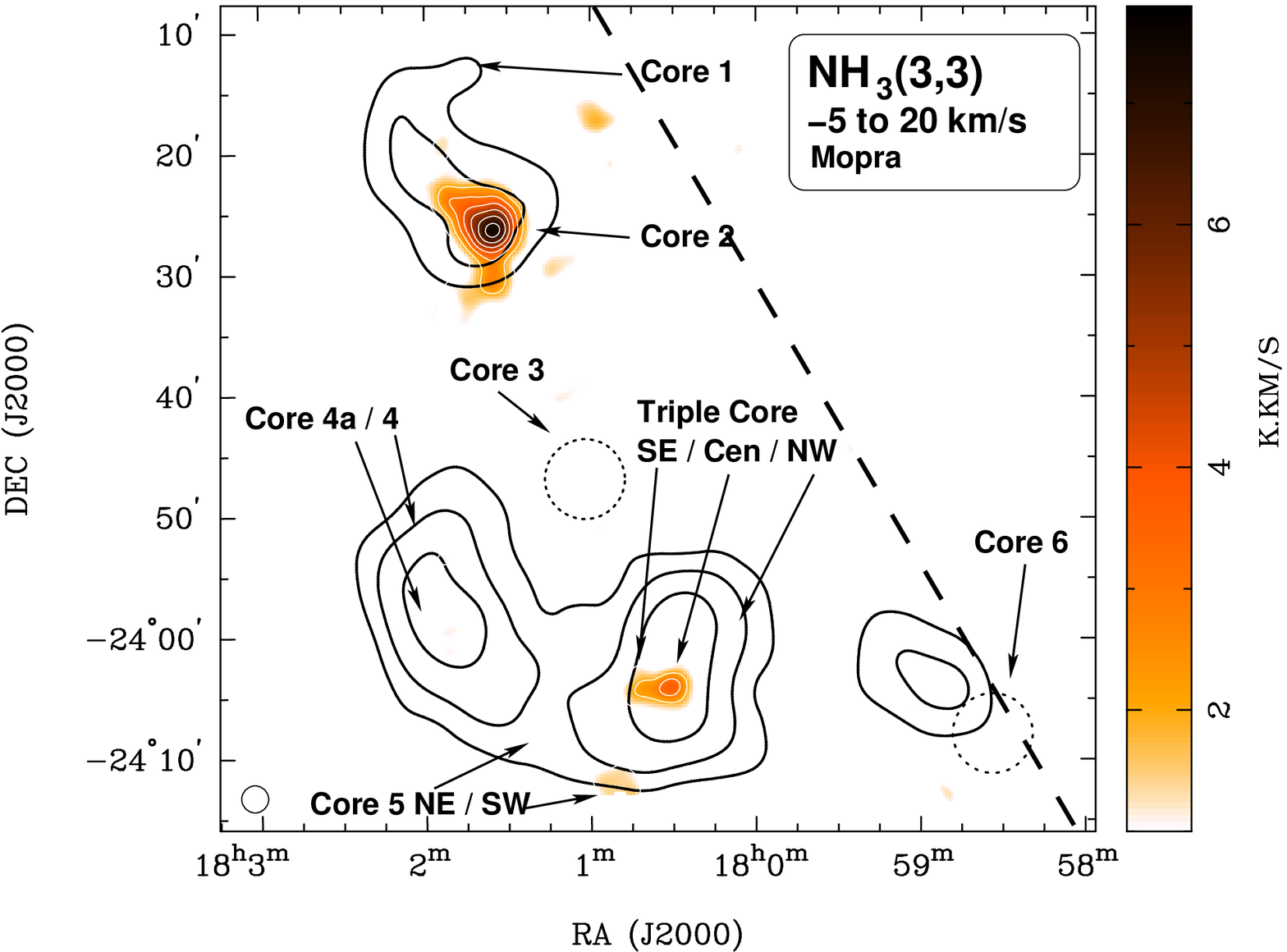}
  \includegraphics[width=0.9\textwidth]{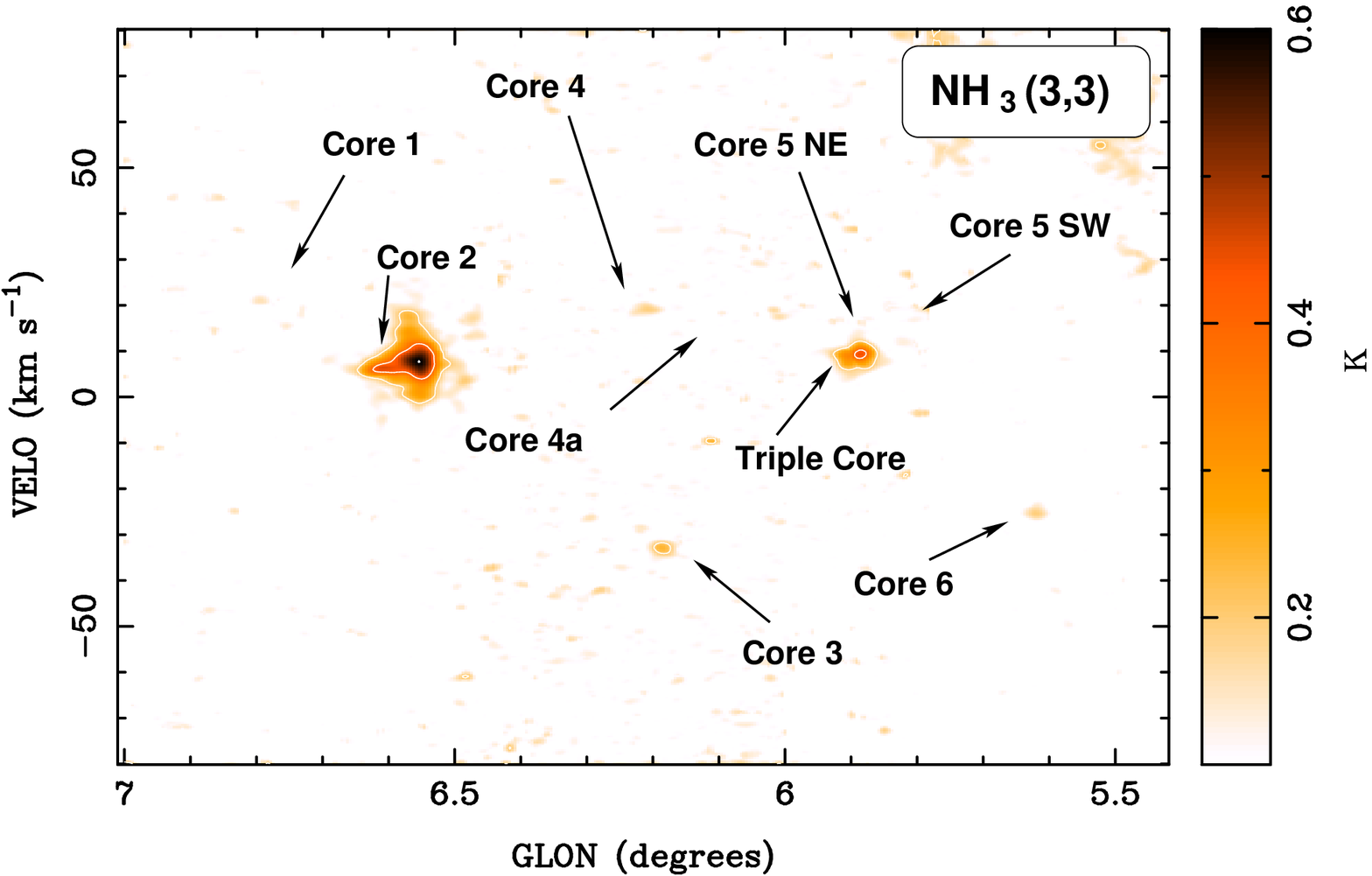}
  \caption[NH$_{3}$\,(3,3) Cores]{As for Figure~\ref{fig:11cores} but showing NH$_{3}$\,(3,3) integrated over -5 to 20\,\kms\, 
    (white contours are 1, 2, 3, 4, 5, 6, 7 K\,\kms) and PV plot (white contours 0.2, 0.4, and 0.6\,K).}
  \label{fig:33cores}
\end{figure*}

Presented in Figures~\ref{fig:11cores},~\ref{fig:22cores} and \ref{fig:33cores} are integrated intensity and position-velocity (PV) plots. The PV plots reveal the velocity-space structure of the NH$_{3}$\,(1,1), (2,2) and (3,3) emission regions or cores in the W28 field. A Hanning smoothing in velocity (width $\sim$6\,\kms) was applied to reduce random fluctuations, and then the Galactic latitude axis was flattened into a single layer.
In this way we show the intrinsic velocity location and width of the gas without confusion.
Based on the PV maps, it is clear that much of the NH$_3$ emission is found in the velocity range $\sim$-5 to $\sim$20\,\kms\, which is quite consistent with the molecular gas
found in CO studies \citep{hess_w28, nanten21,liszt} towards the W28 region. The four satellite lines of NH$_3$\,(1,1) are clearly visible towards most of the cores 
as co-located peaks with 7 and 19\,\kms\, separation from the main line for the inner and outer satellite lines respectively. These satellite lines spread the (1,1) emission over
a wider -20 to 50\,\kms\, range. The intensity maps also in Figures~\ref{fig:11cores},~\ref{fig:22cores} and \ref{fig:33cores} are integrated over \vlsr velocity ranges designed to 
encompass the bulk of the NH$_3$ emission, and show that it is found generally towards the TeV gamma-ray sources, and concentrated into clumps or cores.
For simplicity we label the detected NH$_{3}$ features as Cores~1 to 6 and Triple~Core with some of these containing sub-component clumps, for example, Triple~Core~Central, north west (NW) and 
south east (SE). Table~\ref{tab:corePos} summarises the coordinates of the position-switched spectra towards each core, their relation to our dedicated mapping
and overlapping TeV gamma-ray source.

In mapping data, most cores are seen in both NH$_{3}$\,(1,1) and NH$_{3}$\,(2,2) emission, whereas NH$_{3}$\,(3,3) emission has 
been detected towards the Core~2, Core~5\,SW and the Triple~Core regions. 
In position-switch spectra, NH$_{3}$\,(6,6) is detected towards Core~2 and Triple~Core~Central, while a weak detection of (9,9) is seen towards Core~2 (see online Figures for spectral plots). 
Such transitions are evidence of high gas temperature and potential disruption. 
Core~2 is exceptional in that it represents the well-known NE W28 SNR shock and molecular
cloud interaction region as traced by broad CO emission and 1720\,MHz OH masers towards HESS~J1801-233. A prominent feature here is the 
broadness and relative intensities of NH$_{3}$\,(3,3) and (6,6) compared to the (2,2) and (1,1) transitions. 
The Triple~Core and Core~5 comprise several resolved clumps and are linked to the energetic \hii region G5.89-0.39 at the centre of HESS~J1800-240B. 
Many of the other lines we detect in maps are found towards the Triple~Core and Core~5, namely, bright and extended H$_{2}$O masers, radio recombination lines (RRL), 
and the cyanopolyynes HC$_{3}$N and HC$_{5}$N. Cores~4 and 4a appear to trace additional \hii regions towards HESS~J1800-240A. Core~1 is possibly a star formation site
just north of HESS~J1801-233 and the NE SNR/molecular cloud interaction region. Cores~3 and 6 appear to have quite different velocities at $\sim$-25\,\kms\, to the other
cores and are likely not connected with the molecular gas physically associated with the W28 region.

\subsection{Other 12\,mm Line Detections: H$_{2}$O Masers, Radio Recombination Lines and Cyanopolyynes.}
\label{ssec:OtherLines}

A number of other 12\,mm lines are detected with mapping data over a velocity range consistent with the W28 clouds and our NH$_3$ detections. These are H$_{2}$O masers, the radio recombination lines (RRLs) H62$\alpha$, H65$\alpha$ and H69$\alpha$, and the cyanopolyyne lines HC$_{3}$N(3--2) and HC$_{5}$N(10--9). Maps integrated over the velocity range 5 to 20\,km\,s$^{-1}$, covering the bulk of the detected emission and spectra for all of these detections can be found in the online Figures.
The most prominent detection of H$_{2}$O masers, RRLs and cyanopolyynes is found towards the Triple~Core region, reflecting the strong \hii and star formation activity there. Since the focus of this paper is the dense gas traced by NH$_3$, we only present the measured line parameters for the RRLs, cyanopolyynes and H$_2$O masers in our online appendix. 

In the Triple~Core, H$_{2}$O masers are detected in both the SE and Central regions. The SE emission contains large velocity structure, spread over 100\,\kms.
An additional H$_{2}$O maser is also seen towards Core~1.
According to recent work with 12.2\,GHz methanol masers \citep{breen} an evolutionary sequence for masers associated with high mass star formation regions has been suggested. This evolutionary sequence suggests that the presence of H$_{2}$O masers occurs between $\sim$1.5 to $4.5\times10^{4}$\,yr after high-mass star formation, encompassing the onset of an initial or hypercompact (HC) \hii region forming $\sim$\,2$\times10^{4}$\,yr after the high-mass star formation.

The radio recombination lines trace ionised gas which often exhibit very broad line widths as a result of likely pressure broadening and turbulence 
associated with
ionised gas. The lines we see are no exception to this with several reaching line widths $>40$\,\kms extending to $>100$\,\kms\, in the 
Triple~Core~Central and SE regions. 
Interestingly, the RRL emission appears to be extended and thus may allow probing of the extent to which ionisation is occuring within the southern, central molecular cloud. 
As indicated in Table~\ref{tab:lines} and 
the online appendix we detect 10 RRL transitions in total although only the strongest three transitions (H62$\alpha$, H65$\alpha$, H69$\alpha$) 
are detected in the mapping data.

The strongest cyanopolyyne HC$_{3}$N and HC$_{5}$N emission is found centred on Triple~Core~Central in mapping data but position-switched spectra reveal these molecules towards most
of the other star formation cores (Core~1, 3, 4a, and 6) with a possible weak detection towards Core~2. These are long carbon chain molecules which tend to trace the earlier stages of core evolution 
while NH$_{3}$ tends to become more abundant at later stages \citep{suzuki1992}. However recent work has indicated that HC$_{n}$N($n>3$) can be produced under hot core conditions 
for short periods of time. HC$_{3}$N is produced in large quantities at early times while HC$_{5}$N is created and destroyed within several hundred years, making it a potential chemical clock \citep{chapmann}.

\subsection{CO and Infrared Comparison}
\label{ssec:mwvl}

\begin{figure*}
  \hbox{
    \begin{minipage}{0.6\textwidth}
     \includegraphics[width=\textwidth]{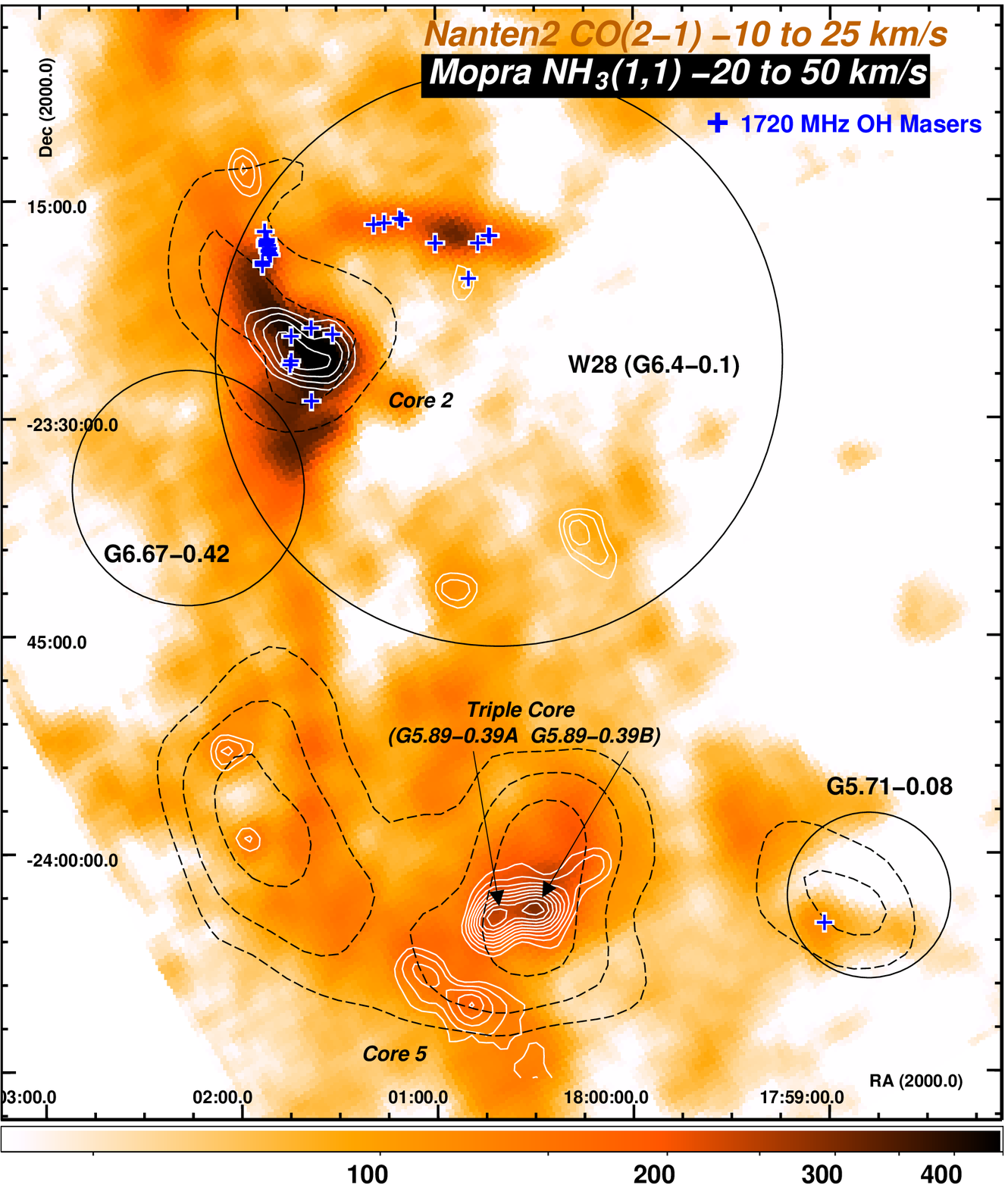}
    \end{minipage}
    \begin{minipage}{0.4\textwidth}
      \centering
      \includegraphics[width=0.95\textwidth]{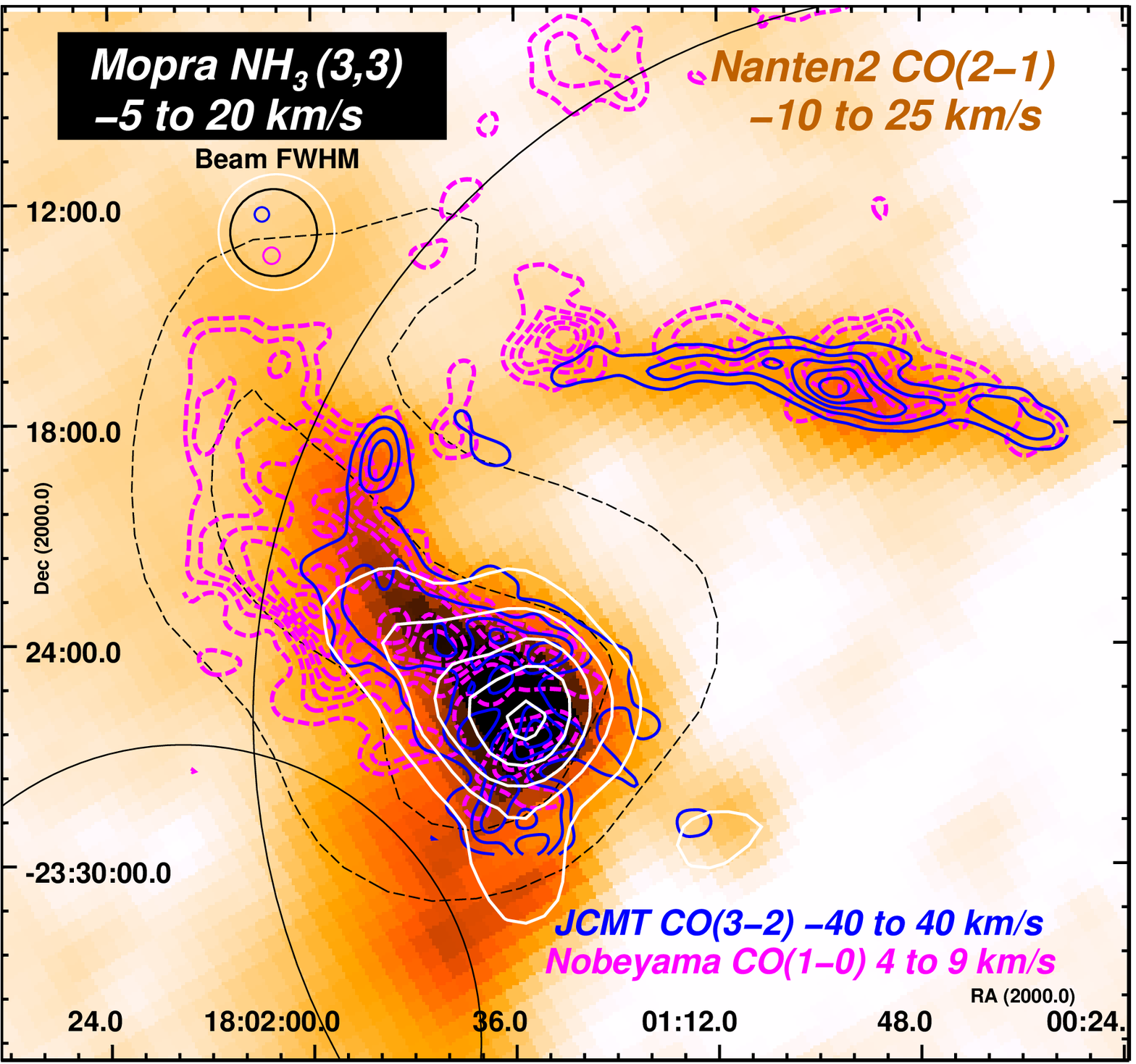}
  \caption[Nanten Comparison]{{\bf Left:} Nanten $^{12}$CO\,(2--1) image [K\,\kms] \citep{nanten21} in log scale, with contours of Mopra NH$_{3}$\,(1,1) (white) and H.E.S.S. TeV gamma-ray significance (black-dashed). 1720\,MHz OH masers from \citet{claussen,hewitt} are also indicated (blue/white $+$). {\bf Above:} Core~2 zoom in linear scale with contours of $^{12}$CO\,(1--0) (magenta-dashed) and $^{12}$CO\,(3--2) (blue) from \citet{arikawa}. SNR diameters for W28 and other SNRs \citep{yusef,brogan2006} are indicated in both panels.
    \label{fig:COdata}}
    \end{minipage}
  }
\end{figure*}

Our 12\,mm line survey adds to the extensive list of molecular cloud observations devoted to the W28 SNR field. The Nanten telescope \citep{nanten} has mapped 
W28 in the $^{12}$CO\,(1--0) \citep{hess_w28} and (2--1) \citep{nanten21} transitions, following on from earlier $^{12}$CO\,(1--0) (e.g. \citealt{arikawa,dame2001,reach,liszt}) 
and  $^{13}$CO\,(1--0) studies \citep{kimkoo2003}.
More recent large-scale surveys of $^{12}$CO\,(2--1) and small scale mapping of the Triple~Core region (G5.89-3.89A and B) in  $^{12}$CO\,(4-3) and 
$^{12}$CO\,(7-6) have
been carried out by Nanten2 \citep{nanten21}. Our Mopra mapping was indeed guided by the Nanten CO results, which provide the most sensitive large-scale look at the molecular 
gas in the region. Figure~\ref{fig:COdata} compares the Nanten2 $^{12}$CO\,(2--1) image with the 
NH$_3$\,(1,1) and (3,3) emission from our Mopra observations, Nobeyama $^{12}$CO\,(1--0) and JCMT $^{12}$CO\,(3--2) observations by \citet{arikawa} as well as 
the TeV gamma-ray emission with H.E.S.S. The $^{12}$CO\,(2--1) emission spatially matches well the brightest three TeV gamma-ray peaks as highlighted previously for 
the $^{12}$CO\,(1--0) emission by \citet{hess_w28}. This match is quite striking towards the Core~2/HESS~J1801-233 region where broad $^{12}$CO\,(3--2) or shocked/disrupted gas tends
to lie inward in the direction of W28 compared to the more quiescent and relatively narrower $^{12}$CO\,(1--0). 
Here, the NH$_3$\,(3,3) emission 
reveals for the first time the dense and disrupted core of the shock-compressed NE molecular cloud. 
Towards the Triple~Core region the Nanten2 $^{12}$CO\,(2--1) emission is resolved into two peaks assocated with the \hii regions G5.89-3.89A and B 
respectively. Core~5 NW and SW are also traced by $^{12}$CO\,(2--1) enhancements, which are also clearly seen in $^{13}$CO\,(1--0) \citep{kimkoo2003}. 
Our Mopra NH$_3$ emission also appears to resolve peaks associated with G5.89-3.89A and B, and Cores~5 NE and SW. 

\begin{figure}
   \centering
   \includegraphics[width=\columnwidth]{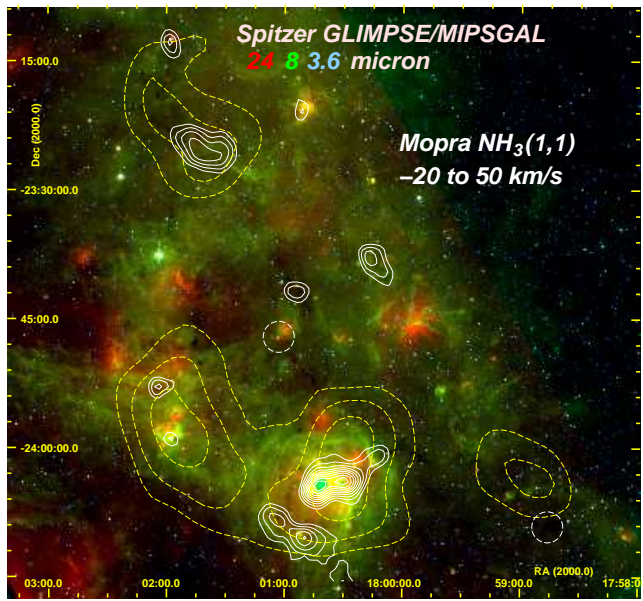}
   \caption[{\it Spitzer} IR data]{{\it Spitzer} GLIMPSE/MIPSGAL three colour (RGB=24/8/3.6\,$\mu$m; MJy/sr in log scale) image with contours from Mopra NH$_{3}$\,(1,1) emission (linear white contours $>$\,1.5\,K) and H.E.S.S. TeV gamma-ray significance (yellow dashed).}
   \label{fig:IRdata}
\end{figure}

Figure~\ref{fig:IRdata} presents the infrared (IR) {\it Spitzer} GLIMPSE and MIPSGAL three-colour (RGB=24/8/3.6\,$\mu$m)
image of the W28 region with Mopra NH$_{3}$\,(1,1) and H.E.S.S. TeV gamma-ray contours. These IR
bands are tracers of polycyclic aromatic hydrocarbons (PAHs) and dust emission, revealing the complexity of the W28 region in hosting 
several star formation and \hii regions in quite likely several different evolutionary stages.
Table~\ref{tab:coreID} summarises the likely counterparts identifed with our detected cores. 

Further discussion of the individual cores 
and comparison of previous molecular line studies toward them can be found later in \S\ref{sec:corediscussion}.
\begin{table}
\centering
\caption{Likely counterparts to the various cores detected. Counterparts have been searched for
  within a 2$^\prime$ radius using the SIMBAD$^\dagger$ astronomical database. Where a star is listed the spectral type is also given.
  \label{tab:coreID}}
  \normalsize
  \begin{tabular}{ll}
  \hline
  Core Name &
  Counterpart Name\\
  \hline
  Core 1			& IRAS 17589-2312$^1$ \\
  Core 2			& W28 SNR/Mol. cloud interaction\\ 
  Core 3			& HD 313632 (B8\,IV)\\
  Core 4 		    	& H{\sc ii} 6.225-0.569$^2$\\
  Core 4a			& H{\sc ii} G6.1-0.6$^3$ / IRAS 17588-2358\\
  Core 5 SW         		& IRAS 17578-2409\\
  Core 5 NE		       	& HD 164194 (B3II/III~C)\\
                                & IRAS 17578-2409$^a$\\
  Triple Core SE		& H{\sc ii} G5.89-0.39A$^4$ / W28-A2\\
  Triple Core Central   	& H{\sc ii} G5.89-0.39B$^4$ \\ 
  Triple Core NW		& V5561 Sgr (M) / IRC-20411$^5$\\ 
  Core 6			& IRAS 17555-2408$^b$\\
  \hline\\[-5mm]
  \multicolumn{2}{l}{\scriptsize 1. \citet{bronfman1996}}\\[-1mm]
  \multicolumn{2}{l}{\scriptsize 2. \citet{lockman}}\\[-1mm]
  \multicolumn{2}{l}{\scriptsize 3. \citet{kuchar}}\\[-1mm]
  \multicolumn{2}{l}{\scriptsize 4. see \citet{kimkoo2001} and references therein.}\\[-1mm]
  \multicolumn{2}{l}{\scriptsize 5. \citet{johnstonH20}.}\\[-1mm]
  \multicolumn{2}{l}{\scriptsize $\dagger$ \url{http://simbad.u-strasbg.fr/simbad/}.}\\[-1mm]
  \multicolumn{2}{l}{\scriptsize $a$. 2.8$^\prime$ distant.}\\[-1mm]
  \multicolumn{2}{l}{\scriptsize $b$. 2.3$^\prime$ distant.}\\[-1mm]
  \hline
\end{tabular}
\end{table}

\section{Analysis of NH$_{3}$ Emission} 
\label{sec:NH3}

We describe here procedures used to estimate the gas parameters such as 
optical depth, temperature, mass, density, and related dynamical information towards the NH$_3$ cores. 
Tables\ref{tab:gas_parameters} (with statistical errors in the online appendix) 
and \ref{tab:mass_dens} summarise these results utilising the 
NH$_3$\,(1,1) and (2,2) spectra for an analysis assuming various core sizes, and additional results treating Core~2, Triple~Core and Core~5 as 
extended clouds well beyond the size of the 2$^\prime$ FWHM beam. We also apply a detailed radiative transfer model to Core~2, given
its apparent high gas temperature and 
non-thermal energy.

\subsection{Line Widths}
\label{ssec:linewidth}
The FWHM of the NH$_{3}$ main line, $\Delta\,v_{1/2}$, is a useful measure of the total
energy associated with the core or clump. Broader lines will result from regions with higher temperatures or
some additional dynamics. The purely Maxwell-Boltzmann thermal line width FWHM $\Delta\,v_{\rm th}$ expected from a gas at 
temperature $T$ is given by 
\begin{equation}
\Delta v_{\rm th} \sim  \sqrt{\frac{8~{\rm ln}(2)~k~T}{m_{\tiny{\textrm{NH}}_{3}}}}\hspace{3 mm}\textrm{[\kms]}
\label{eq:thermal_v}
\end{equation}
where $k$ is Boltzmann's constant and $m_{\tiny{{\rm NH}}_{3}}$ is the mass of the NH$_3$ molecule. For example, a thermal line width of 
$\sim$0.16\,\kms\, is obtained for a temperature of 10\,K, as might be expected in typically cold dense NH$_{3}$ cores \citep{hotownes1983}.
The line FWHM, $\Delta\,v_{1/2}$, of each core was estimated from a Gaussian fit to the central or main peak of the emission with additional Gaussians
to fit each of the four satellite lines, which are generally resolved in the (1,1) transition. This five-Gaussian fit function can be seen applied to 
the Core~1 and Core~2 spectra in Figure~\ref{fig:sample_spectra}. The results in Table~\ref{tab:gas_parameters} show that $\Delta\,v_{1/2}$
for all cores is considerably wider than that expected from purely thermal broadening, suggesting additional non-thermal or kinetic energy which dominates over broadening from the instrumental response, which is considered negligible.

\subsection{Gas Parameters}
\label{ssec:gasparams}

\begin{figure}
  \centering
  \hbox{
    \includegraphics[width=0.55\columnwidth, height=0.25\textheight]{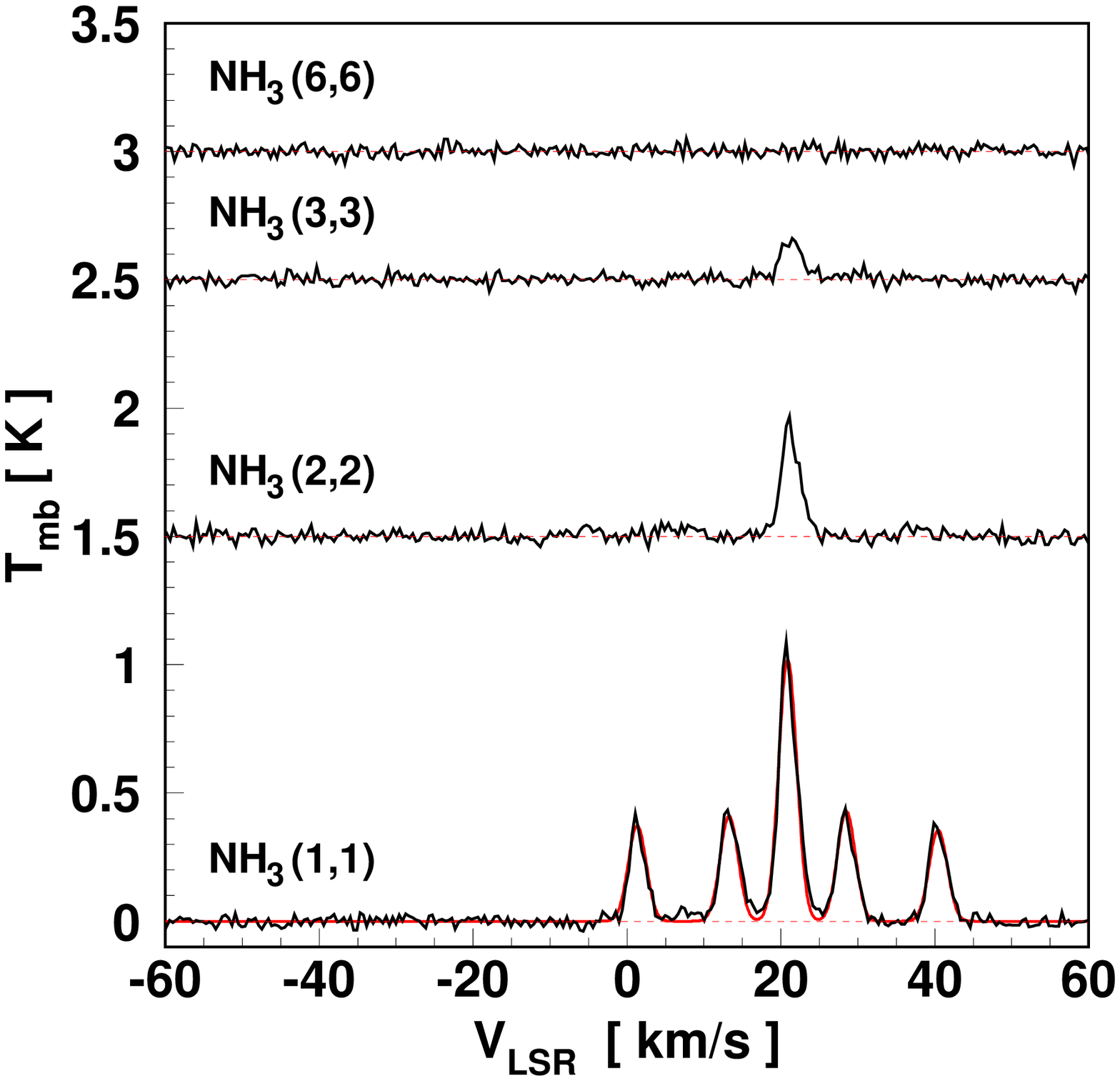}\hspace*{-6mm}
    \includegraphics[width=0.55\columnwidth, height=0.25\textheight]{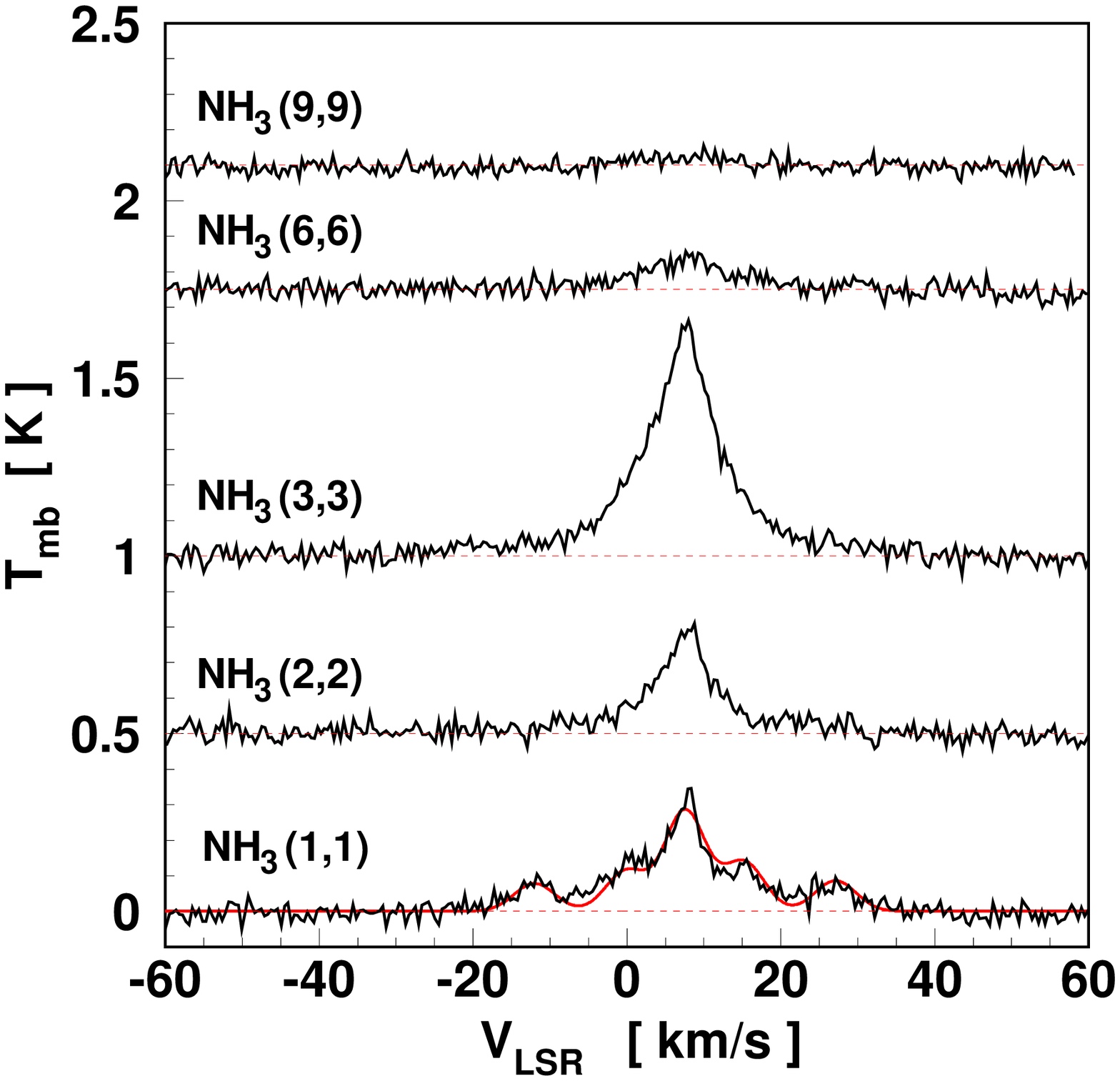}
  }
  \caption{NH$_{3}$\,(n,n) position-switched spectra for Core~1 ({\bf Left}) and Core~2 ({\bf Right}).
    Gaussian fits to the (1,1) and (2,2) spectra used to estimate gas parameters (see Table~\ref{tab:gas_parameters}). The (1,1) fits are shown as red solid lines and  
    the (n,n) spectra for n$\ge$2 are offset by a constant value for clarity.} 
  \label{fig:sample_spectra}
\end{figure}

The (1,1) satellite lines were clearly resolved in all cores except in Core~2 which is instrinsically very broad, leading to blending of the main and satellite lines (see Figure~\ref{fig:sample_spectra}). Using the relative brightness temperatures of the main and satellite peaks, the optical depth can be derived for each ($J,K$) inversion transition by numerically solving Equation~2 of~\citet{barrett}.
A weighted average $\bar{\tau}(J,K,m)$ (with weight given as 1/peak-error according to the Gaussian fit) of the four satellite-derived optical depths is then calculated, and the total optical depth $\tau_{\rm J,K}^{\rm tot}$ of the transition is estimated by accounting for the fraction of intensity in the main line (where $f$(1,1)=0.502, and $f$(2,2)=0.796). Without needing to resolve the (2,2) satellite lines, the main line optical depth of the (1,1) transition can be used to infer the main line optical depth of the (2,2) transition using the method of \citet{unger}. As for the (1,1) transition, a weighted average may be used to derive the total (2,2) optical depth $\tau_{\rm 2,2}^{\rm tot}$.
The line FWHM is of critical consideration when determining the temperature of the gas, as a key assumption in this step is that both the (1,1) and (2,2) emission probe the same volume of gas. The similarity of FWHM for the (1,1) and (2,2) transitions in Table~\ref{tab:gas_parameters} does however suggest that both transitions originate from the same general volume of gas, allowing the rotational temperature to be calculated using the method of \citet{unger}.
Generally the rotational temperature is an underestimate of the kinetic temperature \citep{hotownes1983}, however, the analytical expression of \citet[pg.\,211]{tafalla} has been used to estimate \tkin from \trot. This is believed to be accurate to within 5\% of the real value of \tkin for temperatures in the 5-20\,K range. Importantly, this method for obtaining temperatures is only considered valid for \tkin up to $\sim$40\,K.
To obtain the column density for the NH$_3$ emission, we use the (1,1) transition and follow Equation 9 of \citet{goldsmith_langer}. Since the (1,1) represents only a fraction of the NH$_3$ gas, the total  NH$_3$ column density $N_{\textrm{\tiny{NH}}_{3}}$ can be obtained from the (1,1) column density and the partition function $Q(T)$. In this work we assume contributions up to the $5^{\rm th}$ term 
in the series expansion.

To determine the total mass $M$ and molecular hydrogen number density $n_{\rm H_2}$ (hereafter referred as just density) of the various cores we need to assume the core intrinsic radius $R$ and the abundance ratio 
$\chi_{\textrm{\tiny{NH}}_{3}}$ of NH$_{3}$ to H$_{2}$.
We note that the abundance ratio is known to depend on the gas temperature $T$. Ratios quoted in literature extend from 
$\sim$10$^{-7}$, few~$\times 10^{-8}$ \citep{walmsley1983,ott2005}, down to $\sim 10^{-10}$ \cite{ott2010} in the Large Magellanic Cloud.  
For typical infrared dark clouds (IRDCs) with $T<20$~K an average value 
of $\sim4\times10^{-8}$ is found \citep{pillai}. However with the exception of Core~2 and Core~6, all NH$_3$ cores in the W28 field are identified with star formation
IR sources as indicated in Figure~\ref{fig:IRdata} and Table~\ref{tab:coreID}.
In hot cores, with ($T>100$~K), abundance ratios of 10$^{-6}$ or even as high as 10$^{-5}$ have been suggested (\citealt{walmsley1983,walmsley1992} and 
references therein). We adopt the value 
\begin{equation}
{\large \chi}_{\textrm{\tiny{NH}}_{3}} = N_{\textrm{\tiny{NH}}_{3}}\,/\, N_{\textrm{\tiny{H}}_{2}} = 2\times10^{-8}
\label{abundance}
\end{equation}
following \citet[Table 5.1]{stahler_palla2005} as the temperatures of our detected cores are only slightly higher than those typical 
of IRDCs. 
The total hydrogen mass of the core is then estimated by factoring in the projected area of the emitting region:
\begin{equation}
  M = f K\, 2 m_{\rm H} \, \pi R_{\rm cm}^2 \, N_{\textrm{\tiny{NH}}_{3}}  / ({\large \chi}_{\textrm{\tiny{NH}}_{3}}\,{\rm M_\odot}) \hspace{3 mm}[{\rm M_\odot}]
  \label{eq:mass}
\end{equation}
for the core intrinsic radius expressed in cm units $R_{\rm cm}$ and $m_{\rm H}$ the hydrogen mass (kg). 
Following \citet{mopra_beam} we apply the beam dilution factor $f$, given by $1/f= \ln 2\,(2R/\theta_{\rm mb})^2$, 
and beam coupling coefficient $K=x^2/[1-\exp(-x^2)]$ for $x=\sqrt{1/f}$ in order to account for various source radii.
Therefore the product $fK=1/[1-\exp(-1/f)]$ with $fK\rightarrow 1.0$ for increasing $R$.

The H$_2$ molecular number density $n_{\rm H_2}$ follows by accounting for the emitting volume of emission, assumed to be a sphere with radius $R_{\rm cm}$ in cm.
\begin{equation}
  n_{H_2} = M\,{\rm M_\odot} / (2m_{\rm H}\,4/3\pi R_{\rm cm}^3)\hspace{3 mm}[\rm{cm}^{-3}] 
  \label{eq:density}
\end{equation}
Except for Core~2 and possibly the Triple~Core and Core~5 complexes, the NH$_3$ emission appears to be point-like and thus has an intrinsic diameter upper limit approaching 
the 2$^\prime$ Mopra beam FWHM, or $R_{\rm mb}=0.5\theta_{\rm mb}=0.6$\,pc radius assuming a 2~kpc distance (consistent with the W28 SNR). 
Although $R=0.1$\,pc is typically noted as the intrinsic radius for cold dense cores \citep{hotownes1983}, we assume a source radius of 0.2\,pc as a default.
The estimates for mass and density assuming a point-like source radius of $R=0.2$\,pc are listed in Table~\ref{tab:mass_dens}. In order to allow for varying core radii,
we also include scaling factors for the mass and density resulting from the source radius $R$ dependence in $fK$, source projected area, and volume.

In considering Core~2, Triple~Core and Core~5 as extended regions, we calculate gas parameters from their NH$_3$\,(1,1) and (2,2) spectra 
averaged over elliptical regions and for pixels \tmb $\geq 0.18$\,K. 
The extended mass is estimated as in Eq.~\ref{eq:mass} 
using the column density averaged over the extended region of radius $R_{\rm cm}$ in cm, but with an extra term $\eta_{\rm mb}/\eta_{\rm xb}=0.86$
to account for the extended Mopra beam efficiency $\eta_{\rm xb}=0.7$ appropriate for the NH$_3$ lines \citep{mopra_beam}.
Density then follows from Eq.~\ref{eq:density} as before.
Results, including dimensions of the elliptical regions are given in Tables~\ref{tab:gas_parameters} and \ref{tab:mass_dens}.
Additionally for Core~2, we use the NH$_3$\,(1,1) to (6,6) spectra
(position-switched and mapping) in more detailed radiative transfer modeling discussed in $\S$\ref{ssec:rtmC2} to estimate gas parameters.

Under the assumption that the cores are in gravitational equilibrium with their thermal energy, 
their pure molecular hydrogen virial masses $M_{\rm vir}$ may also be estimated:
\begin{equation}
  M_{\rm vir} = k\, R \, (\Delta v_{1/2})^2 \hspace{3 mm}[\rm{M}_\odot]
  \label{eq:virmass}
\end{equation}
for the source radius $R$ (pc) as before and $\Delta v_{1/2}$ the line FWHM (\,km\,s$^{-1}$). 
The factor $k$ depends on the assumed mass density profile of the core 
$\rho(r)$ with radius $r$. For a Gaussian density profile \citet{protheroe} calculates $k=444$, in contrast to other situations such as a constant density ($k=210$) and
$\rho \propto r^{-2}$ ($k=126$) \citep{maclaren}. Although a Gaussian profile is quite likely, we quote here in Table~\ref{tab:mass_dens} the virial masses bounded by the 
Gaussian and $r^{-2}$ density profiles with a source radius $R$=0.2\,pc. Additionally, an overestimate of the true core line FWHM can result for optically thick
lines. Given the similarity of the NH$_3$\,(1,1) and (2,2) linewidths, and that the (2,2) optical depth is generally less than unity (see Table~\ref{tab:gas_parameters}), 
we use the (2,2) FWHM in the virial mass calculation. The exception is for Core~2 analysis in which case we use the (3,3) linewidth. 
Overall, the virial masses we derive here could be considered upper limits, especially in the case of a Gaussian density profile. Nevertheless, the virial mass
serves as an important guide in understanding the stability of the cores.
 
By far the dominant systematic error in our mass and density estimates arises from the uncertainty in the NH$_3$ abundance ratio ${\large \chi}_{\textrm{\tiny{NH}}_{3}}$. 
Given the range of ratios quoted in literature for cores of similar temperature to ours, we quote
systematic errors of a factor 2 to 5 for ${\large \chi}_{\textrm{\tiny{NH}}_{3}}$, which feed directly into mass and density. Given that most of the core masses we derive
are in agreement with their virial mass range (Table~\ref{tab:mass_dens}), our choice of ${\large \chi}_{\textrm{\tiny{NH}}_{3}}=2\times 10^{-8}$ appears reasonable.

\begin{table*}
\normalsize
\centering
\caption{NH$_3$ gas parameters derived from NH$_3$\,(1,1) and (2,2) spectra. Results for point-like analysis are taken from a single
  position (either from position-switched data or from mapping data in the case of Core~4). Extended source parameters are taken from spectra
  averaged over an elliptical region given below and for pixels satisfying a \tmb $\geq 0.18$\,K masking level in mapping observations.
  Columns from left to right are 
  core name, integrated intensity for (1,1), (2,2) and (3,3),
  rotational temperature \trot, kinetic temperature \tkin, NH$_{3}$\,(1,1,m) main line velocity \vlsr, 
  main line FWHM $\Delta\,v_{1/2}$ for (1,1), (2,2), and (3,3), 
  total column density $N_{\textrm{\tiny{NH}}_{3}}$, and total optical depth {\large $\tau_{\rm 1,1}^{\rm tot}$} and  {\large $\tau_{\rm 2,2}^{\rm tot}$} 
  for (1,1) and (2,2) respectively. Statistical errors are given in the online appendix. 
  \label{tab:gas_parameters}}
\begin{tabular}{lccccccc}
\hline
\multicolumn{1}{c}{Core/Region Name}& 
 \multicolumn{1}{c}{$\int\,\tmb\,dv$}&
  \multicolumn{1}{c}{\trot}& 
   \multicolumn{1}{c}{\tkin$^{\dagger}$}& 
    \multicolumn{1}{c}{\vlsr}& 
     \multicolumn{1}{c}{$\Delta\,v_{1/2}$}& 
      \multicolumn{1}{c}{$N_{\textrm{\tiny{NH}}_{3}}$}&
       \multicolumn{1}{c}{\large{$\tau$}}\\
\multicolumn{1}{c}{  }&  
 \multicolumn{1}{c}{{[K\,\kms]}}&
  \multicolumn{1}{c}{{[K]}}&
   \multicolumn{1}{c}{{[K]}}&
    \multicolumn{1}{c}{{[\kms]}}&
     \multicolumn{1}{c}{{[\kms]}}&
      \multicolumn{1}{c}{{[10$^{13}$\,cm$^{-2}$]}}&
       \multicolumn{1}{c}{  }\\
\multicolumn{1}{c}{  }&
 \multicolumn{1}{c}{\scriptsize (1,1)/(2,2)/(3,3)}&
  \multicolumn{1}{c}{  }&
   \multicolumn{1}{c}{  }&
    \multicolumn{1}{c}{(1,1,m)}&
     \multicolumn{1}{c}{\scriptsize (1,1,m)/(2,2,m)/(3,3,m)} &
      & \multicolumn{1}{c}{\large $\tau_{\rm 1,1}^{\rm tot}$/$\tau_{\rm 2,2}^{\rm tot}$}\\
\hline
\multicolumn{8}{c}{----- Point Source Analysis -----}\\
Core 1                	& 7.6/1.6/0.5  &  16.0	&  18.4  & +21     &  2.8/2.9/3.3  &  36.4   	& 2.7/0.5 \\
Core 2           	& 4.7/2.5/7.4/1.3/0.3$^{\ddagger}$  &  29.9 	&  46.4  & +7  &  6.3/9.7/12.8/15.9/13.8$^{\ddagger}$ &  17.8  & 2.2/1.3 \\
Core 3                	& 8.6/2.5/1.4  &  17.4 	&  20.4  & $-$33   &  3.2/3.3/4.7  &  46.8  	& 3.2/0.7 \\
Core 4                  & 3.8/0.6/--- &  17.9  &  20.7  & +18     &  1.9/1.8/--- &  14.2      & 2.1/0.5 \\
Core 4a               	& 2.1/0.7/0.4  &  20.0 	&  24.9  & +16     &  2.9/3.0/3.4  &  5.5   	& 1.2/0.4 \\
Core 5 SW         	& 6.0/1.1/0.5  &  15.6 	&  17.8  & +16     &  2.3/2.3/3.1  &  25.6  	& 2.3/0.4 \\
Core 5 NE           	& 3.4/0.4/0.2  &  11.6 	&  12.4  & +15     &  1.8/2.0/4.9  &  24.8  	& 3.0/0.2 \\
Triple Core SE      	& 7.8/3.1/1.7  &  20.6	&  25.8  & +9      &  3.4/3.6/4.5  &  35.0  	& 2.8/0.9 \\
Triple Core Cen.        & 8.4/4.2/3.5  &  24.2 	&  32.6  & +9      &  3.9/4.5/5.8  &  23.4  	& 1.5/0.6 \\
Triple Core NW      	& 2.6/0.6/0.4  &  18.6 	&  22.4  & +10     &  2.4/2.1/2.5  &  9.7  	& 2.2/0.5 \\
Core 6                	& 7.9/1.5/0.9  &  14.4 	&  16.1  & $-$26   &  3.0/2.9/3.8  &  67.0	& 4.6/0.6 \\ \hline
\multicolumn{8}{c}{----- Extended Source Analysis -----}\\
Core 2$^1$              & 3.4/1.5/3.5  &  25.2  &  34.8  & +7      &  6.2/8.6/13.5 &  17.0      & 1.6/1.4 \\
Core 5$^2$              & 2.6/0.4/--- &  14.5  &  16.3  & +16     &  3.5/3.9/--- &  14.0      & 1.4/0.4 \\
Triple Core$^3$         & 3.8/1.0/0.8  &  20.1  &  25.0  & +9      &  4.5/4.7/5.2  &  15.0      & 1.2/0.7 \\
\hline\\[-5mm]
\multicolumn{8}{l}{\scriptsize ${\dagger}$ 5\% systematic errors also apply.}\\[-1mm]
\multicolumn{8}{l}{\scriptsize ${\ddagger}$ $\int\,\tmb\,dv$ and $\Delta\,v_{1/2}$ for NH$_3$\,(1,1)/(2,2)/(3,3)/(6,6)/(9,9)}\\[-1mm]
\multicolumn{8}{l}{\scriptsize 1. For ellipse 5.2pc$\times$3.5pc diam. (spherical radius 2.1pc); pos. angle +30$^\circ$; RA 18:01:41 Dec -23:25:06}\\[-1mm]
\multicolumn{8}{l}{\scriptsize 2. For ellipse 7.0pc$\times$2.4pc diam. (spherical radius 2.1pc); pos. angle +30$^\circ$; RA 18:00:49 Dec -24:10:23}\\[-1mm]
\multicolumn{8}{l}{\scriptsize 3. For ellipse 6.6pc$\times$3.1pc diam. (spherical radius 3.2pc); pos. angle -30$^\circ$; RA 18:00:30 Dec -24:03:09}\\[-1mm]
\hline
\end{tabular}
\end{table*}

\subsection{Velocity Dispersion: Core~2 and Triple~Core/Core~5}
\label{ssec:veldisp}
\begin{figure*}
  \includegraphics[width=\textwidth]{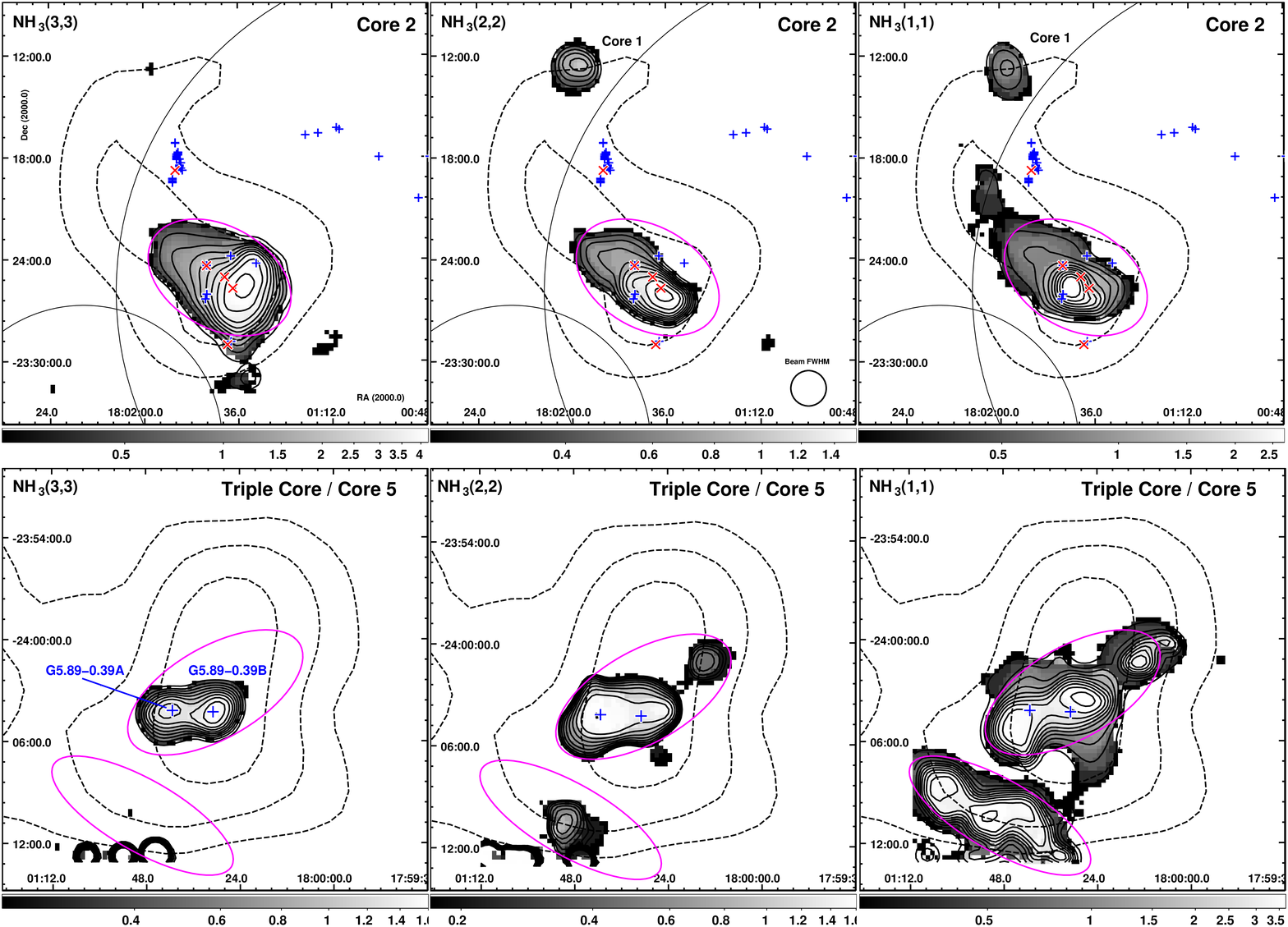}
  \caption{Intensity-weighted velocity dispersion \vrms [\kms] of the NH$_{3}$\,(3,3), (2,2) and (1,1) emission. The (3,3) and (2,2) emission is considered over the -50 to 50\,\kms\, \vlsr range, whilst the (1,1) is considered over the 5 to 15\,\kms\, (Core~2) and  5 to 20\,\kms\, (Triple~Core/Core~5) ranges to avoid the strong satellite lines contaminating the \vrms calculation. Pixel channels are masked below a \tmb value of 0.18\,K ($\sim$3.5\,\trms) threshold. Dashed black contours indicate the H.E.S.S. TeV gamma-ray significance. For Core~2 the blue/white $+$'s indicate the positions of the 1720\,MHz OH masers from \citet{claussen}, and red $\times$'s indicate positions of broad line regions defined by \citet{reach}. The large and small solid circles represent the radio boundaries of the SNRs W28 and G6.67-0.42 respectively. For Triple~Core/Core~5, blue/white $+$'s indicate G5.89-0.39A (\hii) and B (UC-\hii). The Mopra beam 2$^\prime$ FWHM is indicated on the Core~2 NH$_{3}$\,(2,2) image and applies to all other images shown here. The magenta solid ellipses define regions for extended source mass and density estimates in Table~\ref{tab:mass_dens}.The implications of the more central dispersion seen in the Triple Core versus the generally asymmetric dispersion seen in Core~2 are discussed in the text.}
  \label{fig:veldisp}
\end{figure*}

The detection of broad NH$_{3}$ emission primarily towards Core~2 and Triple~Core suggests active disruption of the molecular material. 
The dynamics of the NH$_3$ gas can be probed by looking at the velocity dispersion across a cloud core at each pixel \vrms weighted by its NH$_3$ intensity (\tmb), 
in addition to the position-velocity information (see online appendix).
For each pixel with intensity above a reasonable threshold, in this case 0.18\,K or $\sim$3.5\,\trms,
the velocity dispersion is calculated as:
\begin{equation}
  \vrms = \sqrt {\frac{\int \tmb(v)\,(v-\bar{v})^2\,dv}{\int \tmb(v) dv}} \hspace{3 mm}[\kms]
  \label{eq:veldisp}
\end{equation}
for $\bar{v}\,=\,\int v\,\tmb(v)\,dv/\int \tmb(v)\,dv$\, the intensity-weighted velocity.
Results are presented in Figure~\ref{fig:veldisp} for the Core~2 and the 
Triple~Core/Core~5 regions using the NH$_3$\,(1,1), (2,2) and (3,3) transitions. In the (2,2) and (3,3) transitions a wide -50 to 50\,\kms\, \vlsr velocity 
range encompassing the bulk of the emission was considered. For the (1,1) emission however, the strong satellite lines can contaminate Eq.~\ref{eq:veldisp} and thus a restricted \vlsr range was used for
Core~2 (5 to 15\,\kms) and Triple~Core/Core~5 (5 to 20\,\kms). To further remove fluctuations, data cubes have been Hanning smoothed with velocity width $\sim$2\,\kms. 
For the Core~2 region, the 1720\,MHz OH masers from \citet{claussen} are indicated in Figure~\ref{fig:veldisp}
in order to outline the regions where the W28 SNR shock is interacting directly with the NE molecular cloud. For the Triple~Core/Core~5 region, the 
\hii regions G5.89-0.39A and B are indicated.

While the magnitude of the dispersion varies between the three NH$_{3}$ lines in both cores, the dispersion maps indicate the contrast between Core~2 and the Triple Core/Core~5 region. Core~2 has the most velocity dispersion in the (3,3) line, indicated by both the peak magnitude and the physical area, both of which get progressively smaller in the (2,2) and (1,1) lines. However, the Triple Core/Core~5 region has the most dispersion in the (1,1) line, and progressively less in the (2,2) and (3,3) lines. Interestingly, the spatial location of the peak of the disruption in Core~2 moves radially outward in the direction of the W28 SNR shock, from the western side of the core (in the (3,3) line) toward the eastern side of the core (in the (2,2) and (1,1) lines). For the Triple Core/Core~5 region, the velocity dispersion is always peaked toward the centre of the cores, with no evidence for disruption form an external source. This evidence would suggest that the W28 SNR shock has disrupted Core~2 but not yet reached the Triple Core/Core~5 region.

\subsection{Radiative Transfer Modeling of Core~2}
\label{ssec:rtmC2}

The broad NH$_3$\,(3,3) line 
with  $\Delta v_{1/2}\sim$13\,\kms\, and its high line strength relative to (1,1) and (2,2) from 
Core~2 are somewhat beyond 
those expected from a purely thermal distribution as indicated in $\S$\ref{ssec:linewidth}. Such broad line profiles suggest a large amount of 
energy, in this case non-thermal energy from the SNR shock, has been deposited into the cloud. The velocity 
dispersion NH$_{3}$\,(3,3) image for Core~2 (Figure~\ref{fig:veldisp}) and the many 1720\,MHz OH masers in the region indicating shocked gas
support the notion that the broad NH$_{3}$ is also the result of SNR shock disruption. 
The analysis outlined earlier in $\S$~\ref{ssec:gasparams} 
for the (1,1) and (2,2) emission assumes that it comes from quiescent, 
cold or cool, dense cores. This assumption clearly doesn't hold for the Core~2 region, and as such the mass and density estimates for Core~2 in Table~\ref{tab:mass_dens} are
likely underestimates which would pertain primarily to the cooler gas component. We therefore turn to more detailed radiative transfer modeling  
of the position-switched (1,1), (2,2), (3,3), and (6,6) spectra taken towards the
peak of NH$_3$\,(3,3) emission (Figure~\ref{fig:sample_spectra}) and spectra averaged over the Core~2 region for a first look at the gas parameters towards Core~2.   

The radiative transfer code, MOLLIE\footnote{See \citet{keto1990} for a description of the MOLLIE code, \citet{keto2004} for a description of the line-fitting 
and \citet{keto_zhang2010} and \citet{longmore2010} for recent examples of work using the code}, can deal with arbitrary 3D geometries, but as a first step in 
obtaining the indicative properties of Core~2 in this paper, we modeled the emission as arising from a sphere with a constant temperature, density and non-thermal 
velocity component, taking the NH$_3$ spectra corrected for the 
Mopra aperture main beam efficiency of $\eta_{\rm mb}=0.6$. The $\nhthree$ to H$_2$ abundance ratio was fixed at $2\times10^{-8}$ as in our earlier analyses, and
a source radius of 2.1\,pc (distance 2\,kpc) was chosen based on the extent of the NH$_3$\,(3,3) intensity after a \tmb $\geq0.18$\,K cut.
Models were constructed with H$_2$ densities, temperatures and non-thermal linewidths ranging from $10^3$ to $10^8$\,\cm3, $10$ to $400$\,K and $0.5$ to $40$\,\kms, respectively.  
Radiative transfer modeling was then used to generate synthetic data cubes with a velocity resolution of 0.1\,\kms\, for the $\nhone$, (2,2), (3,3) and (6,6) emission. 
These were then convolved with 2D Gaussian profiles at a spatial scale corresponding to the 2$\arcmin$ FWHM of Mopra. The synthetic spectra at each transition were fit to the observed spectra 
(weighted by the signal-to-noise of 
each transition) and reduced-$\chi^2$ values returned for the goodness-of-fit. Simulated annealing with 10,000 models was used to search through the 3D parameter 
space to minimise $\chi^2$ and find the best-fit model. 
This method is inherently robust against becoming trapped in local, rather than global minima in parameter space. 
However, to determine the robustness of the best-fit model we ran the fitting 20 times with widely separated initial start values and increments. 
For the position-switched spectra, the best-fit model yielded a hydrogen atom number density, temperature and non-thermal linewdith of $10^{3.45}$\,\cm3, 95\,K and 7.4\,\kms, respectively, giving a Core~2 mass of $\sim$2700\,M$_\odot$. For the mapping-averaged spectra, the best-fit model yielded a density, temperature and non-thermal linewdith of $10^{3.12}$\,\cm3, 60\,K and 7.5\,\kms, respectively, giving a Core~2 mass of $\sim$1300\,M$_\odot$.  

\begin{figure}
  \centering
  \includegraphics[width=\columnwidth]{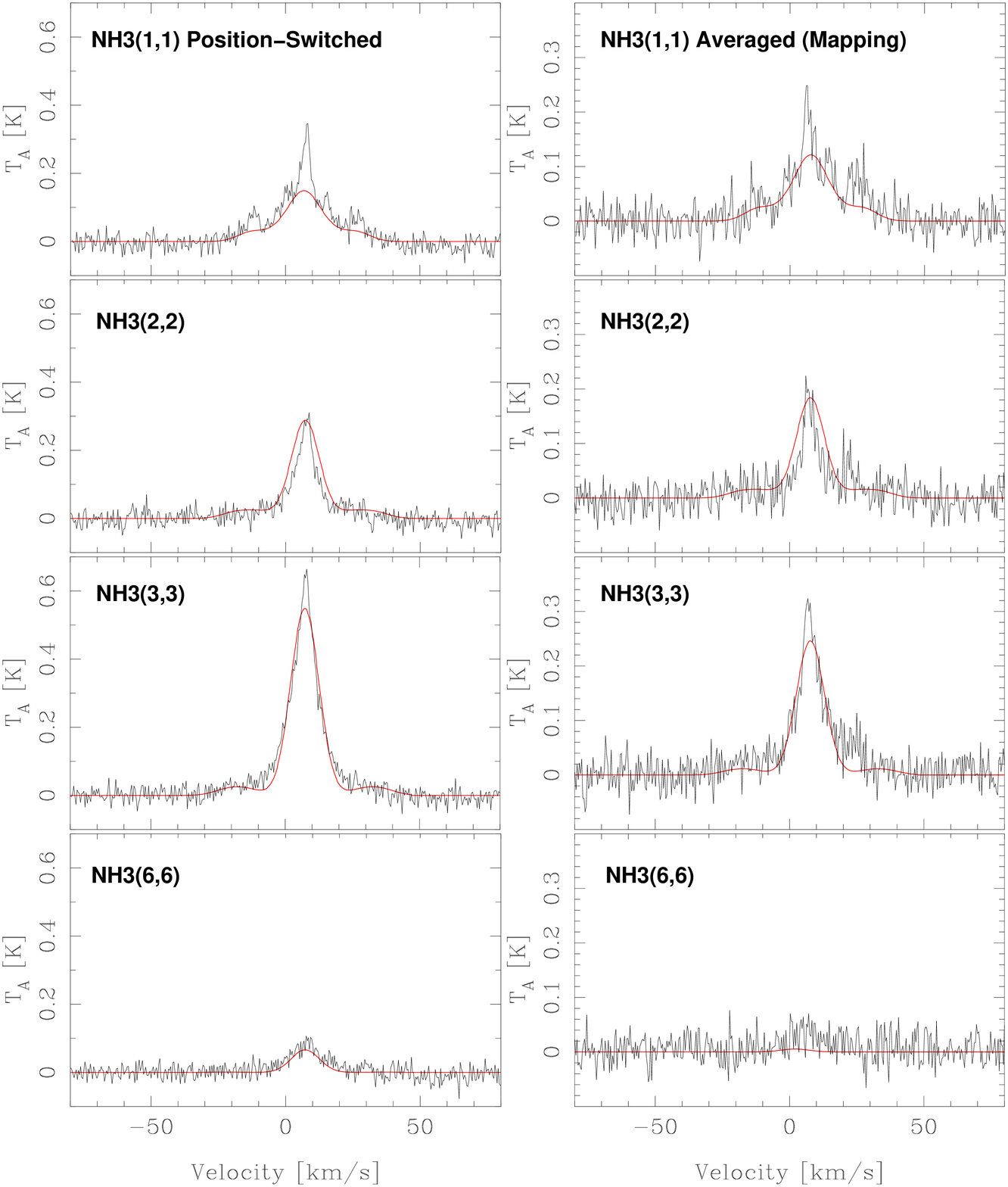}
  \caption{MOLLIE radiative transfer modeling fits to the NH$_3$ spectra from Core~2
    outlined in $\S$\ref{ssec:rtmC2}. {\bf Left-Column:} Fits to position-switched deep spectra
    towards the peak NH$_3$\,(3,3) emission. {\bf Right-Column:} Fits to mapping spectra averaged
    over pixels with \tmb $\geq 0.18$\,K. In all panels the best-fit model
    is overlaid as the red solid line.}
  \label{fig:w28_rt_fit}
\end{figure}

Figure~\ref{fig:w28_rt_fit} shows the NH$_3$\,(1,1), (2,2), (3,3) and (6,6) Mopra spectra from position-switched and averaged-mapping data overlayed with the synthetic 
spectra from the best-fit model. The weakly detected (9,9) spectrum was not included in both cases. Considering the 
simplicity of the model, the synthetic spectra match the position-switched data well. 
Differences between the model and data provide insight into the underlying source structure of 
Core~2. The single non-thermal velocity contribution to the linewidth works well for the higher transitions, but can not account for the narrow linewidth component 
of the NH$_3$\,(1,1) emission from both sets of spectra used.
Similarly, assuming a single temperature for Core~2 underestimates the NH$_3$\,(6,6) emission, especially for the mapping-averaged spectra, which yield a lower density and 
temperature compared to position-switched results as expected. Given these issues, the Core~2 extended masses derived here are likely to be
underestimates and we conservatively quote the lower of the two densities and masses in Table~\ref{tab:mass_dens}. 
More detailed modeling to simultaneously fit the cooler gas traced by the 
narrow linewidth  NH$_3$\,(1,1) emission and the hotter gas traced by the NH$_3$\,(6,6) emission will be the focus of later work. Errors in the fitted quantities 
are not yet estimated due to the often imperfect fits to each spectra, and will also be discussed in later work.

\begin{table*}
\centering
\caption{Mass $M$ and molecular hydrogen number density $n_{\rm H_2}$ estimates (Eqs.~\ref{eq:mass} and \ref{eq:density}) 
  for the various cores from $N_{\textrm{\tiny{NH}}_{3}}$ derived using
  NH$_3$\,(1,1) and (2,2) parameters of Table~\ref{tab:gas_parameters}.
  Point source analysis assumes a spherical emission volume of radius $R=0.2$\,pc (with the effect of source radius scaling indicated) whilst for extended source analysis  
  spectra are averaged over an elliptical region (see Table~\ref{tab:gas_parameters}) using pixels satisfying \tmb $\geq 0.18$\,K.
  The included range of virial masses $M_{\rm vir}$ is bounded by radial density profiles following $r^{-2}$ and Gaussian laws and 
  assumes the NH$_3$\,(2,2) linewidth $\Delta v_{1/2}$ (except for Core~2 which assumes the (3,3) linewidth). 
  The radiative transfer results for Core~2 using the MOLLIE code applied to the NH$_3$\,(1,1) to (6,6) spectra are also given.
  \label{tab:mass_dens}}
\normalsize
\begin{tabular}{lccc}
\hline
Core Name & $M$ & $n_{\rm H_2}$ & $M_{\rm vir}$\\
& [M$_{\odot}$] & [$10^{3}$\,\cm3] & [M$_{\odot}$] \\
\hline
\multicolumn{4}{c}{----- Point Source Analysis ($R = 0.2$\,pc) -----}\\
Core 1			& 	495    	        & 	287.9    & 220 -- 760 \\
Core 2			&       240   		&	140.8    & 4200 -- 14700$\,^\dagger$ \\
Core 3			&	640     	&	370.2    & 280 -- 980 \\
Core 4                  &       190             &       112.3    & 80 -- 290 \\
Core 4a			&	75    		&	43.5     & 230 -- 810 \\
Core 5 SW		&	350   		&	202.5    & 140 -- 480 \\
Core 5 NE		&	335     	&       196.1    & 100 -- 360 \\
Triple Core SE		&	475   		&	276.8    & 330 -- 1170 \\
Triple Core Central	&	320  		&	185.1    & 520 -- 1820 \\
Triple Core NW		&	130   		&	76.7     & 110 -- 400 \\
Core 6			&       910   	        &	530.0    & 220 -- 760 \\ \hline
\multicolumn{3}{l}{\small \underline{Point source mass/density scaling vs. radius $R$ (pc)}} & \\[-0.5mm]
\small $R$ (pc)                                              & \multicolumn{3}{l}{\small 0.10   0.15   0.20   0.25  0.30  0.35  0.40} \\[-0.5mm] 
\small Mass $M(R)\, / \,M(0.2{\rm pc})$                      & \multicolumn{3}{l}{\small 0.97   0.98   1.00   1.02  1.05  1.08  1.12} \\[-0.5mm]
\small Density $n_{\rm H_2}(R)\, / \,n_{\rm H_2}(0.2{\rm pc})$  & \multicolumn{3}{l}{\small 7.77   2.33   1.00   0.52  0.31  0.20  0.14} \\ \hline
\multicolumn{4}{c}{----- Extended Source Analysis -----}\\
Core 2                 &       1600                    & 0.8           & 44400 -- 152600$\,^\dagger$ \\
Core 2 (MOLLIE)        &    $>$1300                    & $>$0.7        & 14500 -- 51200$\,^\ddagger$ \\
Core 5                 &       1300                    & 0.7           & 4000 -- 14200 \\
Triple Core            &       3300                    & 0.5           & 8900 -- 31400  \\
\hline
\multicolumn{4}{l}{\scriptsize $\dagger$ Using $\Delta v_{1/2}$ from the (3,3) emission in Table~\ref{tab:gas_parameters}.}\\[-1mm]
\multicolumn{4}{l}{\scriptsize $\ddagger$ Using $\Delta v_{1/2}$ as the 
non-thermal 
line width from MOLLIE.}\\[-1mm]
\end{tabular}
\end{table*}

\section{Detailed Discussion of Cores}
\label{sec:corediscussion}

\noindent\textit{\bf Core 1:}

Situated at the northern boundary of HESS~J1801-233 in the vicinity of the giant \hii region M20. 
Core~1 exhibits NH$_{3}$\,(1,1), (2,2) and (3,3) emission centred at \vlsr=+21\,\kms\,
with \tkin $\sim18$\,K.
The (1,1) and (2,2) lines are detected in mapping while the (3,3) line is only seen in the deep pointing. CS\,(2--1) was earlier detected at a similar \vlsr value 
found by \citet{bronfman1996}, who first suggested the link to the IR source IRAS~17589-2312. Based on the FIR colour ratio, these authors 
have also suggested IRAS~17589-2312 is indicative of an UC-\hii region. \citet{faundez} estimate a core mass of $M=210$\,M$_{\odot}$, radius $\sim$0.25\,pc
(for 3.8\,kpc distance), and density $n_{\rm H_2}=3.4\times 10^4$\,\cm3 based on their SIMBA 1.2\,mm continuum mapping. Assuming a 0.25\,pc radius, the mass and
density estimates from our observations (500\,M$_{\odot}$ and 15.0$\times 10^4$\,\cm3) are a factor 2 to 4 times higher than those of \citet{faundez}, 
but in general agreement given the systematic uncertainties arising from the NH$_3$ abundance ratio. A similar core radius is also quoted
by \citet{lefloch} who studied this core using a variety of CO, CS, and HCO$^+$ transitions.
They suggest that star formation activity in this core is very recent which could have been triggered by the SNR W28. The H$_2$O maser seen towards this core
has also been discussed \citep{codella}.

\vspace{5mm}

\noindent\textit{\bf Core 2:}

As already described, Core~2 spatially coincides very well with the TeV gamma-ray source HESS~J1801-233 and broad-line CO line 
observations \citep{arikawa,torres,reach,hess_w28,nanten21}. This molecular cloud is known to be shock-disrupted as evidenced by the presence of many 1720\,MHz 
OH masers \citep{frail,claussen}. Further detailed studies of the shocked gas have been carried out in CO, CS and IR H$_2$ lines by \citet{reach}.
Our Mopra observations reveal NH$_{3}$ inversion transitions from (1,1) up to (9,9) with very broad linewidths. 
The strongest emission with $\int\,\tmb\,dv=\,7.4$\,K\,\kms\, is found in the broad $\Delta\,v_{1/2}$=12.8\,\kms\, (3,3) 
transition, which bears a close resemblance to the CO peaks. 
The velocity dispersion image of the (3,3) line in Figure~\ref{fig:veldisp} shows clearly that the cloud disruption originates from the western or 
W28 SNR side, supporting the results of \citet{arikawa} who showed that the broad $^{12}$CO\,(3--2) emission is found preferentially towards the W28 side in contrast
to $^{12}$CO\,(1--0) which extends radially further away. The Nanten2 $^{12}$CO\,(2--1) emission likely traces a mixture of shocked and unshocked gas and detailed velocity 
dispersion studies of this (2--1) emission are currently underway. The lack 
of strong broad-band IR features towards Core~2 (Figure~\ref{fig:IRdata}) suggests that the NH$_{3}$ excitation is not due to star formation processes,
although some fraction of the weak IR emission seen here is due to shocked H$_2$ \citep{neufeld,marquez-lugo}. 
We also note there is a weak HC$_3$N feature seen in the deep pointing which is an indicator of hot gas-phase chemistry.
It is also quite striking that a grouping of the OH masers appear to surround the region where the (3,3) emission is most disrupted
(Figure~\ref{fig:veldisp}), radially away from the W28 SNR. This would be further evidence in support of the W28 SNR as the source of molecular cloud disruption, and
overall would tend to disfavour physical influence from the neighbouring SNR G6.67-0.42, which at present has an unknown distance. The broad molecular line regions
discussed by \citet{reach} also generally cluster towards the broadest NH$_3$ emission (see Figure~\ref{fig:veldisp}).

The broadening of the NH$_3$\,(3,3), (6,6) and probably (9,9) lines are dominated by non-thermal component(s) and we can estimate the additional kinetic 
energy $W_{\rm kin}$ required to achieve this from:
\begin{equation}
  W_{\rm kin} =~ 1/2\,M\,(\Delta v_{\rm kin})^{2}
 \label{eq:kin}
\end{equation}
where $M$ is the mass of the broad-line gas
and $\Delta v_{\rm kin}$ is the FWHM [m\,s$^{-1}$] of the line due to additional non-thermal kinetic processes. Using the
non-thermal line width $\Delta v_{\rm kin}=7.5$\,\kms\, and mass lower limit $M=1300$\,M$_\odot$ from our radiative transfer modeling in \S\ref{ssec:rtmC2}
of mapping-averaged NH$_3$ spectra
we therefore calculate W$_{KE}>0.7\times10^{48}$\,erg. This energy lower limit is within a factor few of the kinetic energy ($\sim\,3\times10^{48}$\,erg) deposited into 
the 2000\,M$_\odot$ of gas traced by shocked $^{12}$CO\,(3--2) from \citealt{arikawa}.

Over the 0 to 12\,\kms\, \vlsr range, for which the  Nanten $^{12}$CO\,(1--0) emission shows excellent overlap with the TeV gamma-ray source HESS~J1801-233,
the mass of the NE molecular cloud is $\sim 2\times 10^{4}$\,M$_{\odot}$. Over a wider 0 to 20\,\kms\, \vlsr range the mass is $\sim 5\times 10^{4}$\,M$_{\odot}$.
Thus, the $>$1300\,M$_\odot$ of extended gas traced by our broad-line NH$_3$ observations represents at least 5\% of the total cloud mass.

\vspace{5mm}

\noindent\textit{\bf Cores 3 and 6:}

Cores~3 and 6 are not found towards any of the H.E.S.S. TeV gamma-ray sources. Their \vlsr values at $\sim$-25\,\kms\, are quite different to the other
Cores in the region. The most likely connection is with the near-3-kpc spiral arm (with heliocentric distance $\sim$2 to 3\,kpc), which has an 
expected value \vlsr $=-53.1+4.16l$\,[\kms] for Galactic longitude $l$ (see e.g. \citealt{dame2008}). 
Core~3 is possibly associated with the B8\,IV spectral type star HD~313632. Our new detection of a H$_2$O maser towards Core~3 may result from the envelope of HD~313632 or
signal the presence of star formation. From the {\em Spitzer} image in Figure~\ref{fig:IRdata} a 24$\mu$m feature is seen towards this core. For Core~6, the IR and radio sources IRAS~17555-2408 and 
PMN~J1758-2405 are found $\sim 2^\prime$ and $\sim 4^\prime$ distant. For both cores, our detection of NH$_3$ could signal some degree of star formation. 
Finally, Core~6 also lies just outside the TeV emission from HESS~J1800-240C and thus is unlikely to be associated. For a 0.2\,pc radius, both of these cores
have masses $\sim$600 to 900\,M$_{\odot}$, and densities $\sim$400 to 500$\times 10^{3}$\,\cm3. 

\vspace{5mm}

\noindent\textit{\bf Cores 4 and 4a}

These cores are located towards the peak of the TeV gamma-ray source HESS~J1800-240A and are likely associated with the \hii regions G6.225-0.569 (Core~4) 
and G6.1-0.6 (Core~4a) 
\citep{lockman,kuchar}. 
Additional counterparts to Core~4a are the IR source, IRAS\,17588-2358, also clearly visible in the {\em Spitzer} image (Figure~\ref{fig:IRdata}), and a 1612\,MHz OH maser \citep{sevenster}.
Our detection of HC$_{3}$N(3--2) towards Core~4a may also suggest it is at an earlier evolutionary stage than Core~4. 
Assuming a 0.2\,pc radius we derived a mass and density of 75 to 200\,M$_{\odot}$ and $n_{\rm H_2}\sim$45 to 100$\times 10^{3}$\,cm$^{-3}$
for Core~4a and Core~4 respectively. 
Despite the reasonable amount of clumpy molecular gas traced by Nanten CO observations (e.g. $\sim 2.5\times 10^{4}$\,M$_{\odot}$ from $^{12}$CO\,(1--0) 
\citealt{hess_w28} assuming a 2~kpc distance), we see no clear indication for extended NH$_3$ emission towards this region. 
We note however this region has only half the Mopra exposure compared to the other regions due to the lack of HOPS overlap.

\vspace{5mm}

\noindent\textit{\bf Triple Core and Core~5}

The Triple~Core represents the most complex of the regions we mapped, and comprises three NH$_3$ peaks aligned in a SE to NW direction all of which are
generally centred on the TeV gamma-ray source HESS~J1800-240B and the very IR-bright and energetic \hii region G5.89-0.39.
Our 12\,mm observations are the largest scale mapping in dense molecular gas tracers so far of this enigmatic region.
G5.89-0.39 actually comprises two active star formation sites and following the nomenclature of \citet{kimkoo2001} they are labeled 
G5.89-0.39A to the east, and G5.89-0.39B about 2$^\prime$ to the west. The Triple~Core~SW NH$_3$ core is associated with 
\hii core G5.89-0.39A, otherwise known as W28-A2 after its strong radio continuum emission. The ring-like features prominently visible in the
{\em Spitzer} 8$\mu$m emission (see Figure~\ref{fig:IRdata}) are centred on G5.89-0.39A suggesting strong PAH molecular excitation from stellar photons.
The Triple~Core~Central NH$_3$ core is associated with the UC-\hii region G5.89-0.39B from which strong H76$\alpha$ RRL appears to 
be centred \citep{kimkoo2001} signalling strong ionisation of the surrounding molecular
gas. The RRL  H62$\alpha$, H65$\alpha$ and H69$\alpha$ emission from our observations also appears prominent towards G5.89-0.39B although in all
three lines the emission is elongated towards G5.89-0.39A. The strongest H$_2$O maser is also seen towards G5.89-0.39B. From the position-switched
observations, strong H$_{2}$O maser emission with complex structures spanning very wide velocity coverage over 100\,\kms\, are detected 
towards G5.89-0.39A and 60\,\kms\, in G5.89-0.39B. 
G5.89-0.39B is responsible for very energetic outflows and is extensively studied in many molecular lines over arcsec to arcmin scales. 
(see e.g. \citealt{harvey,churchwell,gomez,acord,sollins,thompson,klaassen,kimkoo2001,kimkoo2003,hunter}). Previous small-scale NH$_3$ studies are discussed
by \citet{gomez, wood, acord,hunter}.
The Triple~Core~NW NH$_3$ core is found a further 5$^\prime$ distant and may be linked to the M spectral 
type pulsating star V5561\,Sgr or perhaps the natal gas from which this star was born. 

Core~5 appears to straddle the south east quadrant of the 8$\mu$m IR shell or excitation ring of G5.89-0.39A and HESS~J1800-240B, and is resolved into two 
components Core~5 NE and SW. 
Local peaks in CO emission overlapping Core~5 NE and SW are clearly visible (see Figure~\ref{fig:COdata} and
and also \citet{liszt,kimkoo2003}). Core~5~NE is the coldest of the cores detected with \tkin $\sim12$\,K. 
Here we detect only NH$_{3}$\,(1,1) emission in mapping, and only 
very weak (2,2) and (3,3) emission in the deep spectra. It is also one of the few sites where HC$_{5}$N(10--9) is detected.
In Core~5 SW we find \tkin $\sim18$\,K and NH$_3$\,(2,2) and (3,3) emission being stronger than 
in the NE. There is also HC$_{3}$N(3--2) emission when looking at the 5 to 20\,\kms\, velocity range in mapping data.
Overall, assuming a 0.2\,pc core radius, we find our NH$_{3}$\,(1,1) and (2,2) observations trace $\sim$100 to 450\,M$_\odot$ and density 
$n_{\rm H_2}=$80 to 300$\times 10^{3}$\,cm$^{-3}$ for the individual cores in the Triple~Core and Core~5 complexes.
The core masses are in general agreement with their virial masses.

Higher resolution NH$_3$\,(3,3) observations of G5.39-0.39B by \citet{gomez} with the VLA suggest a 0.2\,pc radius molecular envelope around G5.89-0.39B 
(Triple~Core~Central) tracing 
$\sim$30\,M$_\odot$ assuming an abundance ratio ${\large \chi}_{\textrm{\tiny{NH}}_{3}} =10^{-6}$. Using our abundance ratio this mass converts to $\sim$1500\,M$_\odot$,
about a factor 5 larger than our mass estimate using the NH$_3$\,(1,1) and (2,2) emission. This difference may arise from the fact that our analysis is not sensitive
to the slightly broader (3,3) line. In addition \citep{purcellphd} estimates a core mass (for radius 0.15\,pc) using SIMBA 1.2\,mm continuum observations of
360\,M$_\odot$ and derives similar virial masses from the linewidths of N$_2$H$^+$(1--0), H$^{13}$CO\,(1--0), $^{13}$CO\,(1--0) and CH$_3$OH transitions, which are in 
general agreement with our results. 

Assuming the Triple~Core and Core~5 complexes represent extended sources, or at least the superposition of many unresolved point sources,
we derive masses and densities (Table~\ref{tab:mass_dens}) of $\sim$3300 and 1300\,M$_\odot$ respectively with  
$n_{\rm H_2}\sim 0.7\times10^{3}$\,\cm3 from the average NH$_3$\,(1,1) and (2,2) emission. As for Core~2, such estimates do not consider the (3,3)
emission seen towards Triple~Core~Central and SE and are therefore likely to underestimate the true extended mass. 
Nevertheless using these mass estimates the individual NH$_3$ cores represent about 35\% of the extended mass traced by NH$_3$, 
for the two complexes, highlighting their generally clumpy nature.
We also find that the extended mass traced by NH$_3$ represents only small fraction $\sim$5\% of the total cloud mass, 
$\sim 0.8\times10^5$\,M$_\odot$, for the HESS~J1800-240B 
region traced by the Nanten  $^{12}$CO\,(1--0) observations over the \vlsr=0 to 20\,\kms\, range \citep{hess_w28}.

An interesting question is whether or not the W28 SNR shock has reached the southern molecular clouds, as it appears to have done for Core~2 in the NE. 
Of interest therefore is the spatial distribution of any disruption in the molecular clouds associated with the Triple~Core and Core~5 complexes. 
The velocity dispersion image in Figure~\ref{fig:veldisp} clearly shows the broader NH$_3$ gas is found concentrated towards the central star formation cores 
in constrast to Core~2 where the disruption appears to originate more from one side. As a result we would conclude there is no evidence (within the sensitivity
limits of our mapping) for any external disruption of the southern molecular clouds. 

\vspace{5mm}

\section{Summary and Conclusions}
\label{sec:conc}

We have used the Mopra 22m telescope for 12\,mm line mapping over a degree-scale area covering the dense molecular gas towards the W28 SNR field. 
Our aim has been to probe the dense molecular cores of the gas spatially matching the four TeV gamma-ray source observed by the H.E.S.S. telescopes
in this region.
The wide 8\,GHz bandpass of the Mopra telescopes allows to search for a wealth of molecular lines tracing dense gas. For our purpose, the empphasis has
been on the inversion transitions of NH$_3$ which allow relatively robust estimates of gas temperature and optical depth. Our observations combine
data from dedicated scans and those from the the 12\,mm HOPS project and reveal dense
clumpy cores of  NH$_3$ towards most of the molecular cloud complexes overlapping the gamma-ray emission. Additional 12\,mm lines were detected, including the 
cyanopolyynes HC$_{3}$N and HC$_{5}$N, H$_2$O masers, and radio recombination lines, which are prominent towards the UC-\hii 
complex G5.89-0.39. 

The NH$_3$ cores are generally found in regions
of $^{12}$CO peaks and mostly represent sites of star formation at various stages from pre-stellar cores to \hii regions. A standout exception to this
is the shocked molecular cloud on the north east boundary of the W28 SNR where we detect very broad $\Delta\,v_{\rm 1/2}> 10$\,\kms\, NH$_3$ emission 
up to the (9,9) transition with kinetic temperature $>$40\,K. The velocity dispersion of this broad gas further supports the idea that the SNR W28 is the source
of molecular disruption.

Based on analysis of the NH$_3$\,(1,1) and (2,2) transitions we estimate total core masses for pure hydrogen in the range 75 to 900\,M$_{\odot}$ assuming a core radius of 0.2\,pc. 
Molecular hydrogen densities $n_{\rm H_2}$ are in the range 40 to 500$\times 10^3$\,\cm3, however scaling factors for various core sizes included in Table~\ref{tab:mass_dens} indicate that density varies with core radius while mass is more robust to changes in $R$. 
Obviously the estimates for core masses and densities are heavily dependent on the chosen NH$_{3}$ abundance, however, the general agreement between the derived mass and virial mass for the detected cores appears to vindicate our choice of $\chi_{\textrm{\tiny{NH}}_{3}}$.
Optical depths are in the range 1--4 and 0.2--1.2 for the (1,1) and (2,2) transitions respectively. The north east NH$_3$ core (Core~2) is extended and our 
radiative transfer analysis of the broad line components imply a total mass of at least 1300\,M$_\odot$ and density $n_{\rm H_2}>0.7\times 10^3$\,\cm3. 
This represents $>$5\%
of the total north east molecular mass traced by $^{12}$CO\,(1--0) observations. For the southern molecular clouds harbouring \hii regions, the velocity dispersion centres
on the individual cores (Triple~Core SE/Central/NW and Core~5 NE/SW) towards the centre of the molecular clouds. Thus at this stage we find no evidence 
(limited by the $\sim$0.05\,K per channel sensitivity of our mapping) 
for cloud disruption from external sources such as from W28 to the north, and thus it's likely the W28 shock has yet to reach the southern molecular clouds. 
Of course this does not constrain the much faster transport of cosmic-rays from W28 to the southern clouds.

Additionally we note the densities for Core~2, Triple~Core and Core~5 (Table~\ref{tab:mass_dens}) averaged over extended spherical regions are about a factor 10 below the 
critical density for NH$_3$. This would suggest that much of the cores mass is probably contained within a smaller averaged volume than that considered here, 
emphasising their clumpy nature. 
Another possible explanation for the lower densities could be the choice of spherical geometry, especially for the Core~2 region it is plausible that the shock has compressed the core deforming its geometry.

This work is part of our ongoing study into the molecular gas towards the W28 region, and help to further understand internal structures and dynamics of the molecular 
gas in the W28 SNR field. 
Deeper observations in 2010 reaching a $\sim$0.02\,K channel sensitivity in the NH$_{3}$\,(1,1) to (6,6) (including (4,4)) transitions, 
have been made with Mopra in 12\,mm towards the NE (Core~2). 
These deep observations will permit pixel-by-pixel determination of gas parameters such as density, temperature and line width. They will therefore permit a much more detailed probe of the effects of the SNR shock propagating into the Core~2 cloud, using theoretical predictions of the effects of shocks in molecular clouds~\citep{hollenbach_mckee1979,draine1983,hollenbach_mckee1989}.
Additionally, we recently completed a survey in the 7\,mm band to trace the disrupted and shocked gas 
with the SiO\,(1--0) and CS\,(1--0) lines. Finally, we would also add that deeper gamma-ray observations of the 
W28 sources will allow higher resolution gamma-ray imaging approaching the molecular core sizes revealed in this study. 
This, when combined with knowledge of molecular cloud structures on arcminute scales, will
pave the way to probe the diffusion properties of cosmic-rays potentially producing the TeV gamma-ray emission.

\section*{Acknowledgements}
B. N. would like to thank Nigel Maxted for useful discussions and the referee for comments which improved the quality of this work.
This work was supported by an Australian Research Council grant (DP0662810). The Mopra Telescope is part of the Australia Telescope and is funded by the Commonwealth
of Australia for operation as a National Facility managed by CSIRO. The University of New South Wales Mopra Spectrometer Digital Filter Bank used for these Mopra 
observations was provided with support from the Australian Research Council, together with the University of New South Wales, University of Sydney, Monash University
and the CSIRO.

\label{lastpage}

\end{document}